\documentclass[%
 reprint,
 superscriptaddress,
 amsmath,amssymb,
 aps,
 prl,
]{revtex4-2}

\usepackage{graphicx}
\usepackage{dcolumn}
\usepackage{amsmath}
\usepackage{orcidlink}
\usepackage{bm}

\usepackage{xcolor}
\definecolor{apsblue}{RGB}{0,0,139} 
\usepackage{hyperref}
\hypersetup{
    colorlinks=true,
    pdfborder={0 0 0},
    urlcolor=apsblue, 
    linkcolor=apsblue,  
    citecolor=apsblue  
}

\begin{document}

\preprint{APS/123-QED}

\title{Topology of Photonic Time Crystals under Complex Refractive Index Modulation}

\author{Huili Lin\orcidlink{0009-0003-8093-9199}}
\affiliation{\href{https://ror.org/03dz8mb53}{State Key Laboratory of Photonics and Communications}, School of Physics and Astronomy, Shanghai Jiao Tong University, Shanghai 200240, China}

\author{Zhaohui Dong\orcidlink{0000-0002-6388-6879}}
\email{Contact author: zhaohuidong@sjtu.edu.cn}
\affiliation{\href{https://ror.org/03dz8mb53}{State Key Laboratory of Photonics and Communications}, School of Physics and Astronomy, Shanghai Jiao Tong University, Shanghai 200240, China}

\author{Xianfeng Chen\orcidlink{0000-0002-7574-2047}}
\affiliation{\href{https://ror.org/03dz8mb53}{State Key Laboratory of Photonics and Communications}, School of Physics and Astronomy, Shanghai Jiao Tong University, Shanghai 200240, China}
\affiliation{Collaborative Innovation Center of Light Manipulation and Applications, \href{https://ror.org/01wy3h363}{Shandong Normal University}, Jinan 250358, China}
 
\author{Luqi Yuan\orcidlink{0000-0001-9481-0247}}
\email{Contact author: yuanluqi@sjtu.edu.cn}
\affiliation{\href{https://ror.org/03dz8mb53}{State Key Laboratory of Photonics and Communications}, School of Physics and Astronomy, Shanghai Jiao Tong University, Shanghai 200240, China}

\begin{abstract}

Photonic time crystals, constructed by periodic modulation of electromagnetic parameters in time, hold unique topological properties. Here, we explore permittivity and conductivity co-modulated PTCs (PCM PTCs) by dynamically tuning permittivity and conductivity simultaneously of the spatially-uniform materials, and discover their distinctive non-Hermitian effects. We find that such PCM PTCs hold a global shift of the Floquet spectrum along the imaginary quasifrequency axis, which controls field amplification and decay in both momentum bands and band gaps. By further considering the temporal interface consisting of two different PCM PTCs, we can unveil the relationship between the emergence of topological edge states and momentum band inversion by analyzing the eigenstates at exceptional points. Moreover, the characteristic of the topology of PCM PTCs can also be captured from the complex Dirac mass. Our work studies the topology of PTCs in complex modulation regime, offering new opportunities for manipulating topological edge states in the time domain. 
\end{abstract}

\maketitle

Photonic time crystals (PTCs) \cite{morgenthaler1958velocity,asgari2024theory,dong2026photonic}, as a novel class of artificial materials, have emerged as an effective platform for dynamically manipulating light. Such materials usually hold periodic refractive index variations in the temporal dimension while maintaining spatial homogeneity, which induces time-reflections and time-refractions \cite{mendoncca2002time,moussa2023observation,jones2024time}, as well as the resulting momentum bands and bandgaps supporting growing or decaying modes \cite{zurita2009reflection,reyes2015observation,martinez2016temporal,wang2023metasurface,dong2025nonuniform,dong2025extremely}. Owing to these fundamental differences from their spatial counterparts, PTCs attract recent research interest \cite{park2022revealing,wang2025expanding,ni2025topological,sustaeta2025quantum,kiselev2025symmetry,li2026multi,allard2026broadband} and exhibit broad application potentials in areas such as non-resonant tunable lasers \cite{lyubarov2022amplified,li2023stationary,park2025spontaneous,lee2026analogs,huang2026microwave}, topological photonics \cite{lustig2018topological,ma2019band,dong2023band,he2026temporal}, and field enhancement of free electron emission \cite{dikopoltsev2022light,gao2024free}. In particular, in analogy to the topological characterization in 1D spatial photonic crystals, one can define topological invariants for the bands in a PTC, and then also study temporally localized topological edge states at the temporal interface (TI) between two PTCs with different topological phases \cite{lustig2018topological,yang2025topologically,xiong2025observation,jiang2026broadband}. 

While recent studies on PTCs are mostly restricted to lossless media, energy dissipation and gain, the physical origins of complex refractive index, in time-varying dissipative media may greatly change their underlying physics. As a primary route to realizing complex refractive index modulation, complex permittivity modulation in PTCs \cite{wang2018photonic,hayran2021controlling,zhang2025parity,zhou2025band} has been explored, revealing mechanisms including smoothed exceptional-point transition at band edges \cite{wang2018photonic}, the emergence of both momentum and quasienergy bandgaps \cite{zhou2025band}, and tunable temporal parity-time (PT) symmetry phase transitions governed by the ratio between imaginary and real modulation depths \cite{zhang2025parity}. On the other hand, temporal conductivity modulation offers an alternative route to manipulating complex refractive index in the time domain, yet governed by entirely different microscopic mechanisms. Previous study shows that identical temporal refractive index jumps can lead to different wave responses \cite{galiffi2025electrodynamics}, i.e., permittivity modulation ties continuity of electric displacement to charge conservation, whereas conductivity modulation allows continuity of either conduction current or magnetic flux. Owing to these intrinsic boundary differences, conductivity modulation exhibits unique non-Hermitian physical characteristics \cite{harfoush1991scattering,dinc2017synchronized,song2019direction,li2021temporal}, such as broadband lossless non-reciprocity \cite{dinc2017synchronized} and unidirectional thresholdless PT symmetry transition \cite{song2019direction}. Therefore, simultaneous tuning of permittivity and conductivity in PTCs grants full control over the complex refractive index, which might unlock topological physics unattainable in any PTC discussed above.

Here, we study permittivity and conductivity co-modulated PTCs (PCM PTCs) and investigate their band structures, pulse propagation dynamics, and topological properties in the two-dimensional complex refractive index space parameterized by the temporal variations of permittivity $\varepsilon$ and conductivity $\sigma$. By analytically deriving the dispersion relations of PCM PTCs, we reveal how permittivity and conductivity control the shift of the imaginary part of the Floquet quasifrequency $\Omega$, thereby refining the band theory of PCM PTCs. We further study the topological edge state between PCM PTCs with different modulation parameters. We show that the eigenstates at exceptional points (EPs) can signify the occurrence of band inversion, resulting in the emergence of edge states. Moreover, we find that the topological properties of PCM PTCs can be further characterized by studying the complex Dirac mass, which exhibits the fundamentally different characteristic from PTCs with lossless media.

We start with studying the wave propagation in a homogeneous, isotropic and nonmagnetic medium characterized by relative permittivity $\varepsilon$ and conductivity $\sigma$. With real-valued $\varepsilon$ and $\sigma$, the complex refractive index is defined as \cite{li2021temporal}
\begin{equation}
n\equiv\sqrt{\varepsilon+\frac{\sigma}{i\omega\varepsilon_0}}=n'-in'',
\end{equation}
where the real and imaginary parts of the refractive index are $n'=\sqrt{\varepsilon-\tilde{\sigma}^2}$, $n''=\tilde{\sigma}$ with $\tilde{\sigma}\equiv \sigma/2\varepsilon_0k_zc_0$. Here $k_z$ is the wave number, and $c_0=1/\sqrt{\varepsilon_0\mu_0}$ is the speed of light in vacuum, with $\varepsilon_0$ and $\mu_0$ being the vacuum permittivity and permeability. We construct a PCM PTC consisting of two temporal segments, in which the permittivity $\varepsilon(t)$ and conductivity $\sigma(t)$ are periodically modulated in time. Within one period $T$, the complex refractive index $n_1(\varepsilon_1,\sigma_1)$ persists for the first segment of $t_1=T/2$ before it is switched to $n_2(\varepsilon_2,\sigma_2)$ for the second segment $t_2=T/2$, as shown in Fig. 1(a), where $\varepsilon_1=\varepsilon_m+\Delta \varepsilon/2$, $\varepsilon_2=\varepsilon_m-\Delta \varepsilon/2$, $\sigma_1=\sigma_m+\Delta \sigma/2$, and $\sigma_2=\sigma_m-\Delta \sigma/2$. One can obtain the Floquet quasifrequency $\Omega$ as a function of momentum $k$ using the transfer matrix method (see Appendix A in the End Matter). We choose four sets of parameters in the two-dimensional space spanned by $\Delta \varepsilon$ and $\Delta\sigma$ [Fig. 1(b)] and plot the corresponding band structures in Fig. 1(c), where the the blue lines represent the real part of $\Omega T$, and the green lines denote the imaginary part of $\Omega T$. One sees that momentum bandgaps are opened at $\operatorname{Re}(\Omega T)=0, \pm \pi$ in all four cases.

\begin{figure}[htbp]
\includegraphics{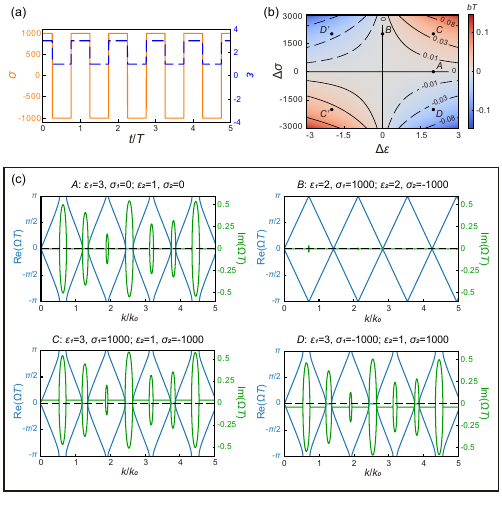}
\caption{\label{fig:epsart} (a) PCM PTC with square-wave modulation of permittivity and conductivity, where $\varepsilon_1=3$, $\sigma_1=1000$, $\varepsilon_2=1$, $\sigma_2=-1000$. (b) Two-dimensional parameter space spanned by $\Delta \varepsilon$ and $\Delta\sigma$, where the background illustrates the global shift of $\operatorname{Im}(\Omega T)$. $\varepsilon_1=\varepsilon_m+\Delta \varepsilon/2$, $\varepsilon_2=\varepsilon_m-\Delta \varepsilon/2$, $\sigma_1=\sigma_m+\Delta \sigma/2$, $\sigma_2=\sigma_m-\Delta \sigma/2$, $\varepsilon_m=2$, $\sigma_m=0$, $T=2\ \mathrm{fs}$. (c) Typical band structures for the four parameter sets (A, B, C, D) marked in (b), where the blue (green) lines denote the real (imaginary) part of $\Omega T$. $k_0=2\pi/Tc_0$.}
\end{figure}

For case A, only the permittivity is modulated, and the conductivity is kept as zero, which has been studied in previous works and is used as the reference here \cite{lustig2018topological}. Now we consider the pure conductivity modulation in case B, where $\sigma_1=-\sigma_2$, and the system is globally lossless and gainless. Nevertheless, momentum bandgaps still open due to the interference between the temporal reflective and refractive waves. We then consider cases C and D with both permittivity and conductivity being modulated. One finds that other than the similar open momentum gaps as those in case A, the imaginary part of $\Omega$ globally shifts upward for case C when $\sigma_1=-\sigma_2=1000$ and $\varepsilon_1>\varepsilon_2$, while it shifts downward for case D when $\sigma_1=-\sigma_2=-1000$. Furthermore, the inequality between $\varepsilon_1$ and $\varepsilon_2$ widens the momentum band gaps and enlarges the difference of $\operatorname{Im}(\Omega T)$ between the two eigenmodes at the gap center.

To explain the above-mentioned global shift of $\operatorname{Im}(\Omega T)$ and explore the influence of permittivity and conductivity on the band structure of PCM PTCs, we derive the analytical expression of the band structure (see Supplemental Materials for detail derivation \cite{suppmat}) as
\begin{equation}
\Omega T=\arccos\left[\frac{1}{2}(X+W)e^{bT}\right]+ibT,
\end{equation}
\vspace{-15pt}
\begin{equation}
b=-\frac{1}{2T\varepsilon_0}\left(\frac{\sigma_1t_1}{\varepsilon_1}+\frac{\sigma_2t_2}{\varepsilon_2}\right),
\end{equation}
where $X+W$ denotes a parameter dependent term on permittivity and conductivity, and $b$ gives the global shift of the imaginary part of the Floquet quasifrequency $\Omega$. We can find that the nonzero value of $b$ arises from the inequality between the effective accumulated gain or loss in each temporal segment, i.e., $\sigma_1 t_1/\varepsilon_1$ and $\sigma_2 t_2/\varepsilon_2$. We plot such global shift of the imaginary eigenvalues under different sets of modulations in Fig. 1(b). One finds that the shift of $\operatorname{Im}(\Omega T)$ increases with both $|\Delta \varepsilon|$ and $|\Delta \sigma|$. 

To investigate the characteristic behavior of light propagation in PCM PTCs, we perform numerical simulations by exciting the band modes and bandgap modes using the finite-difference time-domain (FDTD) method. We set the initial light field as a pulse with its momentum centered at 0.43 $k_0$ (in the band) and 0.71 $k_0$ (in the bandgap), respectively. For \(0 < t < 450\ \mathrm{fs}\), the permittivity \(\varepsilon\) and conductivity $\sigma $ are constant ($\varepsilon=1, \sigma = 0$), and the pulses propagate in free space. At \(t = 450\ \mathrm{fs}\), \(\varepsilon\) and $\sigma$ start to vary periodically in time with the parameters being chosen as those in Fig. 1(c), respectively. After 50 modulation periods, i.e., for \(t > 550\ \mathrm{fs}\), \(\varepsilon\) and $\sigma$ are kept constant again. 

\begin{figure}[htbp]
\includegraphics{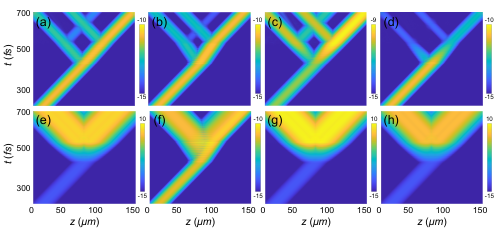}
\caption{\label{fig:epsart} FDTD simulations of pulses propagating in PCM PTCs, illustrating the amplitude of electric displacement field (color bar in log scale). In (a)-(d), the central momentum of the pulses lies in the band, while in (e)-(h), it lies in the bandgap. (a) (e), (b) (f), (c) (g), and (d) (h) correspond to the modulation parameter sets A, B, C and D, respectively. }
\end{figure}

Figs. 2(a)-(d) show simulation results of the displacement field amplitude for a pulse with its central momentum located inside a band of the PCM PTCs. Each pulse splits into two beams at the starting and ending times of temporal modulation, eventually generating four pulses in total. For cases where only permittivity or conductivity is modulated, the Floquet quasifrequency $\Omega$ remains real within the bands, so the pulse amplitude stays nearly constant during propagation. When both permittivity and conductivity are modulated, the nonzero imaginary part of $\Omega$ leads to either growth or decay of the light field, which depends on the sign of the global shift of $\textrm{Im} (\Omega)$.

In contrast, when the momentum of the pulse lies within a bandgap, $\Omega$ is complex in all four cases, leading to exponential growth of field energy over time [Figs. 2(e)-(h)]. Compared with Fig. 2(e), plotting results in the case where only permittivity is modulated, the amplification effect is enhanced in Fig. 2(g) and suppressed in Fig. 2(h) due to the global shift of $\textrm{Im} (\Omega)$. For pure-conductivity modulation case where the bandgap is narrower than the other three cases [Fig. 1(c)], the amplification of the light field is much weaker, as shown in Fig. 2(f). Moreover, since the real part of \(\Omega\) remains constant within the bandgap, the two Floquet modes excited at the onset of temporal modulation exhibit zero group velocity. Consequently, no beam splitting is observed during temporal modulation, and it occurs only at the end of modulation, yielding two pulses finally.

For PTCs with zero conductivity (i.e., real refractive index), a temporal topological edge state appears at TI between two sequential PTCs with different topologies \cite{lustig2018topological}. However, the dynamical behavior of light propagation at TI between two sequential PCM PTCs, or the topology of PCM PTCs, can be very different from the real refractive-index counterpart. We then use the existence of topological edge states as a probe to study the topological phase transition of PCM PTCs.

We begin our analysis with the temporal superlattice $C-C'$, which consists $20T$ in time while TI occurs at $t=10T$. The two constituent PCM PTCs $C$ and $C'$ are illustrated in Fig. 3(a). By choosing $k=0.6073k_0$ within the bandgaps of both sequential PCM PTCs (see Supplemental Material for details \cite{suppmat}), we plot the eigenmode in Fig. 3(b). One sees that the amplitude of the field first increases exponentially and decreases immediately after TI, showing the existence of a temporal topological edge state here. This result indicates that PCM PTCs $C$ and $C'$ share identical band structures but exhibit distinct topologies. 

Similarly, we investigate other temporal superlattices, as illustrated in Figs. 3(c)-(e). For temporal superlattice $C-D$, the two PCM PTCs share the same permittivity modulations but differ in conductivity modulations. The field amplitude increases continuously during temporal evolution, and no topological edge state can be observed. However, for temporal superlattice $D-C'$, the two PCM PTCs only have different permittivity modulations, and a topological edge state can still be obtained by selecting another appropriate momentum $k$ at $0.5994k_0$. Meanwhile, temporal superlattice $D-D'$ also supports topological edge state with the variation of their permittivity $\Delta \varepsilon$ and conductivity $\Delta \sigma$ being opposite in signs.

\begin{figure}[htbp]
\includegraphics{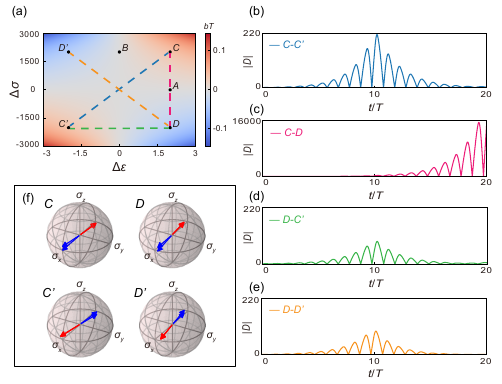}
\caption{\label{fig:epsart} (a) Temporal superlattices with two PCM PTCs cascaded at $t=10T$ (dashed lines in different colors for distinction). (b)-(e) Temporal evolution of the displacement field $|D|$ in the temporal superlattice in (a). (f) EP states in the Bloch sphere for different PCM PTCs, where the blue and red arrows denote eigenstates at the left and right EP, respectively.}
\end{figure}

To understand the phenomena shown above, we use the concept of momentum-band inversion \cite{tong2025observation} to elucidate the underlying physical mechanism. Momentum-band inversion refers to the reversal of the momentum ordering of two bands with distinct eigenstate characters, such that the characteristic originally associated with the lower band becomes associated with the upper band, and vice versa, which hints a possible topological phase transition. Here, we examine the eigenstates at the two exceptional points (EPs) located on the left and right sides of the first momentum bandgap, for simplicity. The two red arrows represent the two eigenstates at the left EP, and the two blue arrows correspond to those at the right EP. Since eigenstates degenerate at an EP, both the blue arrows and red arrows are almost completely overlapped, respectively. It can be seen that momentum-band inversion occurs at the three temporal superlattices ($C-C'$, $D-C'$ and $D-D'$), whereas the EP states remain similar for the other temporal superlattice ($C-D$), which are in accordance with the existence of topological edge states in Figs. 3(b)-(e).

To further explore how the modulation of permittivity and conductivity influences the topological properties of PCM PTCs, we proceed to investigate the momentum-band inversion within the full two-dimensional parameter space spanned by $\Delta\varepsilon$ and $\Delta\sigma$. We find that all EP states mapped on the Bloch sphere lie in the $xOy$ plane (see Supplemental Material \cite{suppmat}). Therefore, we can characterize their orientations by $\varphi_B$, which is defined as the angle between each eigenstate and the $x$-axis. In Fig. 4(a), we plot the angle $\varphi_B$ of the left EP states over the $\Delta\varepsilon-\Delta\sigma$ parameter space. Note that the right EP state has an orientation opposite to that of the left counterpart. One sees that $\varphi_B$ varies by $2\pi$ when $\Delta \varepsilon$ and $\Delta \sigma$ encircling the origin. For any two points collinear with the origin and located on opposite sides, their values of $\varphi_B$ differ by $\pi$, leading to the occurrence of momentum-band inversion. Therefore, the emergence of momentum-band inversion does not have a well-defined boundary in the $\Delta\varepsilon-\Delta\sigma$ space. Instead, it is determined by the relative positions of the two PCM PTCs in the parameter space. 

\begin{figure}[htbp]
\includegraphics{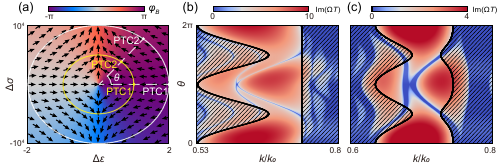}
\caption{\label{fig:epsart} (a) Distribution of $\varphi_B$ for the left EP eigenstates over the $\Delta\varepsilon-\Delta\sigma$ parameter space. The directions of black arrows indicate $\varphi_B$ at the corresponding positions. (b)-(c) The larger imaginary branch of Floquet quasifrequency for the temporal superlattice consisting of two PCM PTCs. Their parameter trajectories are plotted in (a), with the white and yellow lines denoting (b) and (c), respectively. The unshaded region shows the overlap between the bandgaps of the two PCM PTCs.}
\end{figure}

In Fig. 3(d), we see that the emergence of a temporal edge state at superlattice $D-C'$ does not require the angles $\varphi_B$ of the two states to differ strictly by $\pi$. To find the critical angular difference in $\varphi_B$ that supports the existence of edge states, we investigate the band structures of a temporal superlattice consisting of two PCM PTCs. If the imaginary part of the band structure has a zero point at a given momentum $k$, the corresponding eigenstate exhibits neither growth nor decay after it propagates through the temporal superlattice. Instead, its amplitude first increases and then decreases, implying the existence of edge states. Here, the first PCM PTC is fixed with $\Delta\varepsilon=2$ and $\Delta\sigma=0$. For the second one, we follow an elliptical trajectory in the $\Delta\varepsilon-\Delta\sigma$ parameter space [the white line in Fig. 4(a)], with the rotation angle $\theta$ ranging from 0 to $2\pi$. Fig. 4(b) shows the larger imaginary branch of the band structure, where the unshaded region shows the overlap between the bandgaps of the two PCM PTCs. One sees that the zeros of the imaginary eigenvalues only appear for $0.45\pi < \theta < 1.55\pi$, indicating the existence of edge states which corresponds to the difference of the $\varphi_B$ angles of the two PCM PTCs ranges from $0.35\pi$ to $1.65\pi$ (see Supplemental Material for the evolution of eigenstates around the critical angle \cite{suppmat}). Furthermore, different elliptical parameter trajectory for the two PCM PTCs may vary the ranges of the difference between $\varphi_B$ angles. For instance, Fig. 4(c) plots the larger imaginary branch of the band structure but following the yellow trajectory in Fig. 4(a), where the first PTC is fixed with $\Delta\varepsilon=1$ and $\Delta\sigma=0$. In this case, temporal edge states exist when the difference of the two $\varphi_B$ angles ranges from $0.48\pi$ to $1.52\pi$. The robustness of the temporal topological edge state is verified in the Supplemental Material \cite{suppmat}.

The physical origin of these topological edge states can be understood within a Dirac-mass framework \cite{pan2023superluminal}. Under the assumption of weak modulation, the Dirac equation takes the form (see Supplemental Material \cite{suppmat})
\begin{equation}
\small
\left(\frac{1}{c} \sigma_{0}\left(i \frac{\partial}{\partial t}+ \gamma \right)+\sigma_{z}\left(i \frac{\partial}{\partial z}\right)\right) \psi+ iM \psi=0,
\end{equation}
where $\gamma =- {i\mu_0 c^2\Delta\varepsilon\Delta\sigma}/{16\varepsilon_m}$ indicates the global shift of $\textrm{Im} (\Omega)$, $ M=(\mu_0 c\Delta\sigma /8)\sigma_x - ({\omega_m\Delta\varepsilon}/{16c\varepsilon_m})\sigma_y $, and $\sigma_i \ (i=x,y,z)$ denotes Pauli matrices. $iM\psi$ corresponds to the imaginary mass term, which can be written in the complex Dirac mass form
\begin{equation}
m =  m_1 + i m_2 = \frac{\mu_0 c\Delta\sigma}{8}+i\frac{\Delta\varepsilon\omega_m}{16c\varepsilon_m}.
\end{equation}
Meanwhile, $m$ can be expressed as $m = m_0 e^{i\phi}$, where $m_0 = \sqrt{m_1^2 + m_2^2}$ and $\phi = \tan^{-1}(m_2/m_1)$ denote the amplitude and phase of the complex Dirac mass, respectively. We emphasize that the complex number $m$ is a compact expression of the vectorial mass operator $M$, which should not be confused with a genuine complex scalar Dirac mass. Moreover, it is fundamentally different from the complex Dirac mass found in spatial photonic crystals \cite{lu2018topological,gao2020dirac,bi2015unidirectional}, which leads to the opening of frequency bandgaps. In previous studies \cite{ren2025observation,tong2025observation,li2025topological}, it has been noted that the momentum-gap topology is characterized by the sign of Dirac mass $m$. A system composed of two PTCs with opposite Dirac mass signs supports topological edge state at TI. In our work, the modulation parameters $\Delta \varepsilon$ and $\Delta \sigma$ control the topology of PCM PTCs by tuning the phase of the complex Dirac mass. A zero-energy topological mode occurs only when the relative mass phase satisfies $\Delta \phi =\pi$, as the ones in the temporal superlattice $C-C'$ and $D-D'$ in Fig. 3. For adjacent-quadrant interfaces, such as $D-C'$ and $C-D$, temporally localized edge modes may appear, but they are generally displaced from zero energy for $\Delta \phi \neq \pi$, resulting in an energy splitting [see Fig. 4(b)].

In summary, we consider PTCs under complex refractive index modulation and study relevant physical features including their band structures, dynamical evolution, and topological properties. We analytically derive the Floquet dispersion relation and reveal the global shift of $\operatorname{Im}(\Omega T)$. Moreover, we show that temporal edge states emerge when the difference between the $\varphi_B$ angle of two PCM PTCs falls in a certain range. The complex Dirac mass further reveals that the modulation parameters $\Delta\varepsilon$ and $\Delta\sigma$ control the topology of PCM PTCs. This work provides a new insight into the topology of PCM PTCs, extends the band theory of PTCs into the complex parameter space, and offers a feasible route to control temporal topological edge states via combined permittivity and conductivity modulation. 

Our findings can be further extended to higher-dimensional cases, such as two-dimensional topological edge states in photonic space-time crystals \cite{sharabi2022spatiotemporal,segal2025two,zhang2026topological} under complex refractive index modulation. Experimentally, the time-varying systems have been demonstrated in ultracold atom lattices \cite{dong2024quantum}, synthetic lattice platform \cite{yu2025comprehensive} with fiber loops \cite{ye2023reconfigurable,yu2024dirac,qin2024observation,feis2025space,ren2025observation,he2025observing} and dynamically modulated transmission lines \cite{reyes2015observation,xiong2025observation,lee2026analogs}. These existing platforms pave the way towards the future experimental realization of the PCM PTCs proposed in this work.

\bigskip

\acknowledgments
\textit{Acknowledgments}---The research was supported by National Key R\&D Program of China (No. 2023YFA1407200), National Natural Science Foundation of China (12192252).

\appendix
\section{End Matter}
\renewcommand{\theequation}{A\arabic{equation}}
\setcounter{equation}{0}

\textit{Appendix A: The transfer matrix method}---We consider two waves propagating in opposite directions along the $z$ direction with the same wave number $k_z$, and write the total transverse electric and magnetic field components as
\begin{equation}
E_x(z,t)=e^{-ik_zz}\left[E^+(t)+E^-(t)\right]+\text{c.c.};
\end{equation}
\begin{equation}
\eta_0H_y(z,t)=e^{-ik_zz}\left[nE^+(t)-n^*E^-(t)\right]+\text{c.c.},
\end{equation}
with the vacuum impedance $\eta_0=\sqrt{\mu_0/\varepsilon_0}$ , complex frequency $\omega=k_zc_0/n$, and time-dependent amplitudes of the forward (backward) waves $E^+(t)=E_+e^{i\omega t}$ ($E^-(t)=E_-^*e^{-i\omega^* t}$). Here $E_+$ ($E_-^*$) is the complex amplitude constant of forward (backward) waves at the initial time. When a wave propagates in medium with $n_1$ during the $i$-th period from time $t_i$ to $t_i'=t_i+t_1$, the field amplitudes $\bm{\psi}_E(t)\equiv[E^+(t), E^-(t)]^T$ vary as $\bm{\psi}_E(t_i')=P_1( t_1)\bm{\psi}_E(t_i)$, where the propagation matrix is defined as
\begin{equation}
P_1(t_1)=\begin{pmatrix}e^{i\omega_1t_{1}}&0\\0&e^{-i\omega_1^*t_{1}}\end{pmatrix},
\end{equation}
with the complex frequency $\omega_1=k_zc_0/n_1$. At the temporal boundary $t_i'$, the complex refractive index changes abruptly from $n_1$ to $n_2$. Based on the continuity of electromagnetic fields at $t_i'$, the field amplitudes satisfy $\bm{\psi}_E(t_i'^+)=Q_{12}\bm{\psi}_E(t_i'^-)$, with the transfer matrix \cite{li2021temporal}
\begin{equation}
Q_{12}=\begin{pmatrix}
\frac{n_1(n_1^*+n_2)}{n_2(n_2+n_2^*)}&\frac{n_1^*(n_1-n_2)}{n_2(n_2+n_2^*)}\\
\frac{n_1(n_1^*-n_2^*)}{n_2^*(n_2+n_2^*)}&\frac{n_1^*(n_1+n_2^*)}{n_2^*(n_2+n_2^*)}
\end{pmatrix}.
\end{equation}
The propagation matrix $P_2(t_2)$ for the second temporal segment and the transfer matrix $Q_{21}$ at the temporal boundary where $n_2$ changes abruptly to $n_1$ can be written at similar forms. We therefore obtain the transfer matrix $M$ over one period as
\begin{equation}
M=Q_{21}P_2(t_2)Q_{12}P_1(t_1).
\end{equation}
According to the Floquet theorem, we can write the eigenvalue equation $M\bm{D}(t)=e^{-i\Omega T}\bm{D}(t)$ and then solve the Floquet quasifrequency $\Omega$ as a function of the momentum $k$.


\begin{thebibliography}{60}%
\makeatletter
\providecommand \@ifxundefined [1]{%
 \@ifx{#1\undefined}
}%
\providecommand \@ifnum [1]{%
 \ifnum #1\expandafter \@firstoftwo
 \else \expandafter \@secondoftwo
 \fi
}%
\providecommand \@ifx [1]{%
 \ifx #1\expandafter \@firstoftwo
 \else \expandafter \@secondoftwo
 \fi
}%
\providecommand \natexlab [1]{#1}%
\providecommand \enquote  [1]{``#1''}%
\providecommand \bibnamefont  [1]{#1}%
\providecommand \bibfnamefont [1]{#1}%
\providecommand \citenamefont [1]{#1}%
\providecommand \href@noop [0]{\@secondoftwo}%
\providecommand \href [0]{\begingroup \@sanitize@url \@href}%
\providecommand \@href[1]{\@@startlink{#1}\@@href}%
\providecommand \@@href[1]{\endgroup#1\@@endlink}%
\providecommand \@sanitize@url [0]{\catcode `\\12\catcode `\$12\catcode `\&12\catcode `\#12\catcode `\^12\catcode `\_12\catcode `\%12\relax}%
\providecommand \@@startlink[1]{}%
\providecommand \@@endlink[0]{}%
\providecommand \url  [0]{\begingroup\@sanitize@url \@url }%
\providecommand \@url [1]{\endgroup\@href {#1}{\urlprefix }}%
\providecommand \urlprefix  [0]{URL }%
\providecommand \Eprint [0]{\href }%
\providecommand \doibase [0]{https://doi.org/}%
\providecommand \selectlanguage [0]{\@gobble}%
\providecommand \bibinfo  [0]{\@secondoftwo}%
\providecommand \bibfield  [0]{\@secondoftwo}%
\providecommand \translation [1]{[#1]}%
\providecommand \BibitemOpen [0]{}%
\providecommand \bibitemStop [0]{}%
\providecommand \bibitemNoStop [0]{.\EOS\space}%
\providecommand \EOS [0]{\spacefactor3000\relax}%
\providecommand \BibitemShut  [1]{\csname bibitem#1\endcsname}%
\let\auto@bib@innerbib\@empty
\bibitem [{\citenamefont {Morgenthaler}(1958)}]{morgenthaler1958velocity}%
  \BibitemOpen
  \bibfield  {author} {\bibinfo {author} {\bibfnamefont {F.~R.}\ \bibnamefont {Morgenthaler}},\ }\bibfield  {title} {\bibinfo {title} {Velocity modulation of electromagnetic waves},\ }\href@noop {} {\bibfield  {journal} {\bibinfo  {journal} {IRE Transactions on microwave theory and techniques}\ }\textbf {\bibinfo {volume} {6}},\ \bibinfo {pages} {167} (\bibinfo {year} {1958})}\BibitemShut {NoStop}%
\bibitem [{\citenamefont {Asgari}\ \emph {et~al.}(2024)\citenamefont {Asgari}, \citenamefont {Garg}, \citenamefont {Wang}, \citenamefont {Mirmoosa}, \citenamefont {Rockstuhl},\ and\ \citenamefont {Asadchy}}]{asgari2024theory}%
  \BibitemOpen
  \bibfield  {author} {\bibinfo {author} {\bibfnamefont {M.~M.}\ \bibnamefont {Asgari}}, \bibinfo {author} {\bibfnamefont {P.}~\bibnamefont {Garg}}, \bibinfo {author} {\bibfnamefont {X.}~\bibnamefont {Wang}}, \bibinfo {author} {\bibfnamefont {M.~S.}\ \bibnamefont {Mirmoosa}}, \bibinfo {author} {\bibfnamefont {C.}~\bibnamefont {Rockstuhl}},\ and\ \bibinfo {author} {\bibfnamefont {V.}~\bibnamefont {Asadchy}},\ }\bibfield  {title} {\bibinfo {title} {Theory and applications of photonic time crystals: a tutorial},\ }\href@noop {} {\bibfield  {journal} {\bibinfo  {journal} {Advances in optics and photonics}\ }\textbf {\bibinfo {volume} {16}},\ \bibinfo {pages} {958} (\bibinfo {year} {2024})}\BibitemShut {NoStop}%
\bibitem [{\citenamefont {Dong}\ and\ \citenamefont {Yuan}(2026)}]{dong2026photonic}%
  \BibitemOpen
  \bibfield  {author} {\bibinfo {author} {\bibfnamefont {Z.}~\bibnamefont {Dong}}\ and\ \bibinfo {author} {\bibfnamefont {L.}~\bibnamefont {Yuan}},\ }\bibfield  {title} {\bibinfo {title} {Photonic time crystals: Unlocking novel light-matter interaction},\ }\href@noop {} {\bibfield  {journal} {\bibinfo  {journal} {SCIENCE CHINA Physics, Mechanics \& Astronomy}\ }\textbf {\bibinfo {volume} {69}} (\bibinfo {year} {2026})}\BibitemShut {NoStop}%
\bibitem [{\citenamefont {Mendon{\c{c}}a}\ and\ \citenamefont {Shukla}(2002)}]{mendoncca2002time}%
  \BibitemOpen
  \bibfield  {author} {\bibinfo {author} {\bibfnamefont {J.}~\bibnamefont {Mendon{\c{c}}a}}\ and\ \bibinfo {author} {\bibfnamefont {P.}~\bibnamefont {Shukla}},\ }\bibfield  {title} {\bibinfo {title} {Time refraction and time reflection: two basic concepts},\ }\href@noop {} {\bibfield  {journal} {\bibinfo  {journal} {Physica Scripta}\ }\textbf {\bibinfo {volume} {65}},\ \bibinfo {pages} {160} (\bibinfo {year} {2002})}\BibitemShut {NoStop}%
\bibitem [{\citenamefont {Moussa}\ \emph {et~al.}(2023)\citenamefont {Moussa}, \citenamefont {Xu}, \citenamefont {Yin}, \citenamefont {Galiffi}, \citenamefont {Ra’di},\ and\ \citenamefont {Al{\`u}}}]{moussa2023observation}%
  \BibitemOpen
  \bibfield  {author} {\bibinfo {author} {\bibfnamefont {H.}~\bibnamefont {Moussa}}, \bibinfo {author} {\bibfnamefont {G.}~\bibnamefont {Xu}}, \bibinfo {author} {\bibfnamefont {S.}~\bibnamefont {Yin}}, \bibinfo {author} {\bibfnamefont {E.}~\bibnamefont {Galiffi}}, \bibinfo {author} {\bibfnamefont {Y.}~\bibnamefont {Ra’di}},\ and\ \bibinfo {author} {\bibfnamefont {A.}~\bibnamefont {Al{\`u}}},\ }\bibfield  {title} {\bibinfo {title} {Observation of temporal reflection and broadband frequency translation at photonic time interfaces},\ }\href@noop {} {\bibfield  {journal} {\bibinfo  {journal} {Nature Physics}\ }\textbf {\bibinfo {volume} {19}},\ \bibinfo {pages} {863} (\bibinfo {year} {2023})}\BibitemShut {NoStop}%
\bibitem [{\citenamefont {Jones}\ \emph {et~al.}(2024)\citenamefont {Jones}, \citenamefont {Kildishev}, \citenamefont {Segev},\ and\ \citenamefont {Peroulis}}]{jones2024time}%
  \BibitemOpen
  \bibfield  {author} {\bibinfo {author} {\bibfnamefont {T.~R.}\ \bibnamefont {Jones}}, \bibinfo {author} {\bibfnamefont {A.~V.}\ \bibnamefont {Kildishev}}, \bibinfo {author} {\bibfnamefont {M.}~\bibnamefont {Segev}},\ and\ \bibinfo {author} {\bibfnamefont {D.}~\bibnamefont {Peroulis}},\ }\bibfield  {title} {\bibinfo {title} {Time-reflection of microwaves by a fast optically-controlled time-boundary},\ }\href@noop {} {\bibfield  {journal} {\bibinfo  {journal} {Nature Communications}\ }\textbf {\bibinfo {volume} {15}},\ \bibinfo {pages} {6786} (\bibinfo {year} {2024})}\BibitemShut {NoStop}%
\bibitem [{\citenamefont {Zurita-S{\'a}nchez}\ \emph {et~al.}(2009)\citenamefont {Zurita-S{\'a}nchez}, \citenamefont {Halevi},\ and\ \citenamefont {Cervantes-Gonz{\'a}lez}}]{zurita2009reflection}%
  \BibitemOpen
  \bibfield  {author} {\bibinfo {author} {\bibfnamefont {J.~R.}\ \bibnamefont {Zurita-S{\'a}nchez}}, \bibinfo {author} {\bibfnamefont {P.}~\bibnamefont {Halevi}},\ and\ \bibinfo {author} {\bibfnamefont {J.~C.}\ \bibnamefont {Cervantes-Gonz{\'a}lez}},\ }\bibfield  {title} {\bibinfo {title} {Reflection and transmission of a wave incident on a slab with a time-periodic dielectric function $\varepsilon$ (t)},\ }\href@noop {} {\bibfield  {journal} {\bibinfo  {journal} {Physical Review A—Atomic, Molecular, and Optical Physics}\ }\textbf {\bibinfo {volume} {79}},\ \bibinfo {pages} {053821} (\bibinfo {year} {2009})}\BibitemShut {NoStop}%
\bibitem [{\citenamefont {Reyes-Ayona}\ and\ \citenamefont {Halevi}(2015)}]{reyes2015observation}%
  \BibitemOpen
  \bibfield  {author} {\bibinfo {author} {\bibfnamefont {J.}~\bibnamefont {Reyes-Ayona}}\ and\ \bibinfo {author} {\bibfnamefont {P.}~\bibnamefont {Halevi}},\ }\bibfield  {title} {\bibinfo {title} {Observation of genuine wave vector (k or $\beta$) gap in a dynamic transmission line and temporal photonic crystals},\ }\href@noop {} {\bibfield  {journal} {\bibinfo  {journal} {Applied Physics Letters}\ }\textbf {\bibinfo {volume} {107}} (\bibinfo {year} {2015})}\BibitemShut {NoStop}%
\bibitem [{\citenamefont {Mart{\'\i}nez-Romero}\ \emph {et~al.}(2016)\citenamefont {Mart{\'\i}nez-Romero}, \citenamefont {Becerra-Fuentes},\ and\ \citenamefont {Halevi}}]{martinez2016temporal}%
  \BibitemOpen
  \bibfield  {author} {\bibinfo {author} {\bibfnamefont {J.~S.}\ \bibnamefont {Mart{\'\i}nez-Romero}}, \bibinfo {author} {\bibfnamefont {O.}~\bibnamefont {Becerra-Fuentes}},\ and\ \bibinfo {author} {\bibfnamefont {P.}~\bibnamefont {Halevi}},\ }\bibfield  {title} {\bibinfo {title} {Temporal photonic crystals with modulations of both permittivity and permeability},\ }\href@noop {} {\bibfield  {journal} {\bibinfo  {journal} {Physical Review A}\ }\textbf {\bibinfo {volume} {93}},\ \bibinfo {pages} {063813} (\bibinfo {year} {2016})}\BibitemShut {NoStop}%
\bibitem [{\citenamefont {Wang}\ \emph {et~al.}(2023)\citenamefont {Wang}, \citenamefont {Mirmoosa}, \citenamefont {Asadchy}, \citenamefont {Rockstuhl}, \citenamefont {Fan},\ and\ \citenamefont {Tretyakov}}]{wang2023metasurface}%
  \BibitemOpen
  \bibfield  {author} {\bibinfo {author} {\bibfnamefont {X.}~\bibnamefont {Wang}}, \bibinfo {author} {\bibfnamefont {M.~S.}\ \bibnamefont {Mirmoosa}}, \bibinfo {author} {\bibfnamefont {V.~S.}\ \bibnamefont {Asadchy}}, \bibinfo {author} {\bibfnamefont {C.}~\bibnamefont {Rockstuhl}}, \bibinfo {author} {\bibfnamefont {S.}~\bibnamefont {Fan}},\ and\ \bibinfo {author} {\bibfnamefont {S.~A.}\ \bibnamefont {Tretyakov}},\ }\bibfield  {title} {\bibinfo {title} {Metasurface-based realization of photonic time crystals},\ }\href@noop {} {\bibfield  {journal} {\bibinfo  {journal} {Science advances}\ }\textbf {\bibinfo {volume} {9}},\ \bibinfo {pages} {eadg7541} (\bibinfo {year} {2023})}\BibitemShut {NoStop}%
\bibitem [{\citenamefont {Dong}\ \emph {et~al.}(2025{\natexlab{a}})\citenamefont {Dong}, \citenamefont {Zhang}, \citenamefont {He}, \citenamefont {Li},\ and\ \citenamefont {Xu}}]{dong2025nonuniform}%
  \BibitemOpen
  \bibfield  {author} {\bibinfo {author} {\bibfnamefont {J.}~\bibnamefont {Dong}}, \bibinfo {author} {\bibfnamefont {S.}~\bibnamefont {Zhang}}, \bibinfo {author} {\bibfnamefont {H.}~\bibnamefont {He}}, \bibinfo {author} {\bibfnamefont {H.}~\bibnamefont {Li}},\ and\ \bibinfo {author} {\bibfnamefont {J.}~\bibnamefont {Xu}},\ }\bibfield  {title} {\bibinfo {title} {Nonuniform wave momentum band gap in biaxial anisotropic photonic time crystals},\ }\href@noop {} {\bibfield  {journal} {\bibinfo  {journal} {Physical Review Letters}\ }\textbf {\bibinfo {volume} {134}},\ \bibinfo {pages} {063801} (\bibinfo {year} {2025}{\natexlab{a}})}\BibitemShut {NoStop}%
\bibitem [{\citenamefont {Dong}\ \emph {et~al.}(2025{\natexlab{b}})\citenamefont {Dong}, \citenamefont {Chen},\ and\ \citenamefont {Yuan}}]{dong2025extremely}%
  \BibitemOpen
  \bibfield  {author} {\bibinfo {author} {\bibfnamefont {Z.}~\bibnamefont {Dong}}, \bibinfo {author} {\bibfnamefont {X.}~\bibnamefont {Chen}},\ and\ \bibinfo {author} {\bibfnamefont {L.}~\bibnamefont {Yuan}},\ }\bibfield  {title} {\bibinfo {title} {Extremely narrow band in moir{\'e} photonic time crystal},\ }\href@noop {} {\bibfield  {journal} {\bibinfo  {journal} {Physical Review Letters}\ }\textbf {\bibinfo {volume} {135}},\ \bibinfo {pages} {033803} (\bibinfo {year} {2025}{\natexlab{b}})}\BibitemShut {NoStop}%
\bibitem [{\citenamefont {Park}\ \emph {et~al.}(2022)\citenamefont {Park}, \citenamefont {Cho}, \citenamefont {Lee}, \citenamefont {Lee}, \citenamefont {Lee}, \citenamefont {Park}, \citenamefont {Ryu}, \citenamefont {Park}, \citenamefont {Jeon},\ and\ \citenamefont {Min}}]{park2022revealing}%
  \BibitemOpen
  \bibfield  {author} {\bibinfo {author} {\bibfnamefont {J.}~\bibnamefont {Park}}, \bibinfo {author} {\bibfnamefont {H.}~\bibnamefont {Cho}}, \bibinfo {author} {\bibfnamefont {S.}~\bibnamefont {Lee}}, \bibinfo {author} {\bibfnamefont {K.}~\bibnamefont {Lee}}, \bibinfo {author} {\bibfnamefont {K.}~\bibnamefont {Lee}}, \bibinfo {author} {\bibfnamefont {H.~C.}\ \bibnamefont {Park}}, \bibinfo {author} {\bibfnamefont {J.-W.}\ \bibnamefont {Ryu}}, \bibinfo {author} {\bibfnamefont {N.}~\bibnamefont {Park}}, \bibinfo {author} {\bibfnamefont {S.}~\bibnamefont {Jeon}},\ and\ \bibinfo {author} {\bibfnamefont {B.}~\bibnamefont {Min}},\ }\bibfield  {title} {\bibinfo {title} {Revealing non-hermitian band structure of photonic floquet media},\ }\href@noop {} {\bibfield  {journal} {\bibinfo  {journal} {Science advances}\ }\textbf {\bibinfo {volume} {8}},\ \bibinfo {pages} {eabo6220} (\bibinfo {year} {2022})}\BibitemShut {NoStop}%
\bibitem [{\citenamefont {Wang}\ \emph {et~al.}(2025)\citenamefont {Wang}, \citenamefont {Garg}, \citenamefont {Mirmoosa}, \citenamefont {Lamprianidis}, \citenamefont {Rockstuhl},\ and\ \citenamefont {Asadchy}}]{wang2025expanding}%
  \BibitemOpen
  \bibfield  {author} {\bibinfo {author} {\bibfnamefont {X.}~\bibnamefont {Wang}}, \bibinfo {author} {\bibfnamefont {P.}~\bibnamefont {Garg}}, \bibinfo {author} {\bibfnamefont {M.}~\bibnamefont {Mirmoosa}}, \bibinfo {author} {\bibfnamefont {A.}~\bibnamefont {Lamprianidis}}, \bibinfo {author} {\bibfnamefont {C.}~\bibnamefont {Rockstuhl}},\ and\ \bibinfo {author} {\bibfnamefont {V.}~\bibnamefont {Asadchy}},\ }\bibfield  {title} {\bibinfo {title} {Expanding momentum bandgaps in photonic time crystals through resonances},\ }\href@noop {} {\bibfield  {journal} {\bibinfo  {journal} {Nature Photonics}\ }\textbf {\bibinfo {volume} {19}},\ \bibinfo {pages} {149} (\bibinfo {year} {2025})}\BibitemShut {NoStop}%
\bibitem [{\citenamefont {Ni}\ \emph {et~al.}(2025)\citenamefont {Ni}, \citenamefont {Yin}, \citenamefont {Li},\ and\ \citenamefont {Al{\`u}}}]{ni2025topological}%
  \BibitemOpen
  \bibfield  {author} {\bibinfo {author} {\bibfnamefont {X.}~\bibnamefont {Ni}}, \bibinfo {author} {\bibfnamefont {S.}~\bibnamefont {Yin}}, \bibinfo {author} {\bibfnamefont {H.}~\bibnamefont {Li}},\ and\ \bibinfo {author} {\bibfnamefont {A.}~\bibnamefont {Al{\`u}}},\ }\bibfield  {title} {\bibinfo {title} {Topological wave phenomena in photonic time quasicrystals},\ }\href@noop {} {\bibfield  {journal} {\bibinfo  {journal} {Physical Review B}\ }\textbf {\bibinfo {volume} {111}},\ \bibinfo {pages} {125421} (\bibinfo {year} {2025})}\BibitemShut {NoStop}%
\bibitem [{\citenamefont {Sustaeta-Osuna}\ \emph {et~al.}(2025)\citenamefont {Sustaeta-Osuna}, \citenamefont {Garc{\'\i}a-Vidal},\ and\ \citenamefont {Huidobro}}]{sustaeta2025quantum}%
  \BibitemOpen
  \bibfield  {author} {\bibinfo {author} {\bibfnamefont {J.~E.}\ \bibnamefont {Sustaeta-Osuna}}, \bibinfo {author} {\bibfnamefont {F.~J.}\ \bibnamefont {Garc{\'\i}a-Vidal}},\ and\ \bibinfo {author} {\bibfnamefont {P.}~\bibnamefont {Huidobro}},\ }\bibfield  {title} {\bibinfo {title} {Quantum theory of photon pair creation in photonic time crystals},\ }\href@noop {} {\bibfield  {journal} {\bibinfo  {journal} {ACS photonics}\ }\textbf {\bibinfo {volume} {12}},\ \bibinfo {pages} {1873} (\bibinfo {year} {2025})}\BibitemShut {NoStop}%
\bibitem [{\citenamefont {Kiselev}\ and\ \citenamefont {Pan}(2025)}]{kiselev2025symmetry}%
  \BibitemOpen
  \bibfield  {author} {\bibinfo {author} {\bibfnamefont {E.~I.}\ \bibnamefont {Kiselev}}\ and\ \bibinfo {author} {\bibfnamefont {Y.}~\bibnamefont {Pan}},\ }\bibfield  {title} {\bibinfo {title} {Symmetry breaking and spatiotemporal pattern formation in photonic time crystals},\ }\href@noop {} {\bibfield  {journal} {\bibinfo  {journal} {Physical Review A}\ }\textbf {\bibinfo {volume} {111}},\ \bibinfo {pages} {053509} (\bibinfo {year} {2025})}\BibitemShut {NoStop}%
\bibitem [{\citenamefont {Li}\ \emph {et~al.}(2026)\citenamefont {Li}, \citenamefont {Mirmoosa}, \citenamefont {Asadchy},\ and\ \citenamefont {Wang}}]{li2026multi}%
  \BibitemOpen
  \bibfield  {author} {\bibinfo {author} {\bibfnamefont {Z.}~\bibnamefont {Li}}, \bibinfo {author} {\bibfnamefont {M.}~\bibnamefont {Mirmoosa}}, \bibinfo {author} {\bibfnamefont {V.}~\bibnamefont {Asadchy}},\ and\ \bibinfo {author} {\bibfnamefont {X.}~\bibnamefont {Wang}},\ }\bibfield  {title} {\bibinfo {title} {Multi-mode photonic time crystals based on time-modulated metasurface waveguides},\ }\href@noop {} {\bibfield  {journal} {\bibinfo  {journal} {arXiv preprint arXiv:2605.14268}\ } (\bibinfo {year} {2026})}\BibitemShut {NoStop}%
\bibitem [{\citenamefont {Allard}\ \emph {et~al.}(2026)\citenamefont {Allard}, \citenamefont {Sustaeta-Osuna}, \citenamefont {Garc{\'\i}a-Vidal},\ and\ \citenamefont {Huidobro}}]{allard2026broadband}%
  \BibitemOpen
  \bibfield  {author} {\bibinfo {author} {\bibfnamefont {T.~F.}\ \bibnamefont {Allard}}, \bibinfo {author} {\bibfnamefont {J.~E.}\ \bibnamefont {Sustaeta-Osuna}}, \bibinfo {author} {\bibfnamefont {F.~J.}\ \bibnamefont {Garc{\'\i}a-Vidal}},\ and\ \bibinfo {author} {\bibfnamefont {P.~A.}\ \bibnamefont {Huidobro}},\ }\bibfield  {title} {\bibinfo {title} {Broadband dipole absorption in dispersive photonic time crystals},\ }\href@noop {} {\bibfield  {journal} {\bibinfo  {journal} {Physical Review Letters}\ }\textbf {\bibinfo {volume} {136}},\ \bibinfo {pages} {106903} (\bibinfo {year} {2026})}\BibitemShut {NoStop}%
\bibitem [{\citenamefont {Lyubarov}\ \emph {et~al.}(2022)\citenamefont {Lyubarov}, \citenamefont {Lumer}, \citenamefont {Dikopoltsev}, \citenamefont {Lustig}, \citenamefont {Sharabi},\ and\ \citenamefont {Segev}}]{lyubarov2022amplified}%
  \BibitemOpen
  \bibfield  {author} {\bibinfo {author} {\bibfnamefont {M.}~\bibnamefont {Lyubarov}}, \bibinfo {author} {\bibfnamefont {Y.}~\bibnamefont {Lumer}}, \bibinfo {author} {\bibfnamefont {A.}~\bibnamefont {Dikopoltsev}}, \bibinfo {author} {\bibfnamefont {E.}~\bibnamefont {Lustig}}, \bibinfo {author} {\bibfnamefont {Y.}~\bibnamefont {Sharabi}},\ and\ \bibinfo {author} {\bibfnamefont {M.}~\bibnamefont {Segev}},\ }\bibfield  {title} {\bibinfo {title} {Amplified emission and lasing in photonic time crystals},\ }\href@noop {} {\bibfield  {journal} {\bibinfo  {journal} {Science}\ }\textbf {\bibinfo {volume} {377}},\ \bibinfo {pages} {425} (\bibinfo {year} {2022})}\BibitemShut {NoStop}%
\bibitem [{\citenamefont {Li}\ \emph {et~al.}(2023)\citenamefont {Li}, \citenamefont {Yin}, \citenamefont {He}, \citenamefont {Xu}, \citenamefont {Al{\`u}},\ and\ \citenamefont {Shapiro}}]{li2023stationary}%
  \BibitemOpen
  \bibfield  {author} {\bibinfo {author} {\bibfnamefont {H.}~\bibnamefont {Li}}, \bibinfo {author} {\bibfnamefont {S.}~\bibnamefont {Yin}}, \bibinfo {author} {\bibfnamefont {H.}~\bibnamefont {He}}, \bibinfo {author} {\bibfnamefont {J.}~\bibnamefont {Xu}}, \bibinfo {author} {\bibfnamefont {A.}~\bibnamefont {Al{\`u}}},\ and\ \bibinfo {author} {\bibfnamefont {B.}~\bibnamefont {Shapiro}},\ }\bibfield  {title} {\bibinfo {title} {Stationary charge radiation in anisotropic photonic time crystals},\ }\href@noop {} {\bibfield  {journal} {\bibinfo  {journal} {Physical Review Letters}\ }\textbf {\bibinfo {volume} {130}},\ \bibinfo {pages} {093803} (\bibinfo {year} {2023})}\BibitemShut {NoStop}%
\bibitem [{\citenamefont {Park}\ \emph {et~al.}(2025)\citenamefont {Park}, \citenamefont {Lee}, \citenamefont {Zhang}, \citenamefont {Park}, \citenamefont {Ryu}, \citenamefont {Cho}, \citenamefont {Lee}, \citenamefont {Zhang}, \citenamefont {Park}, \citenamefont {Jeon} \emph {et~al.}}]{park2025spontaneous}%
  \BibitemOpen
  \bibfield  {author} {\bibinfo {author} {\bibfnamefont {J.}~\bibnamefont {Park}}, \bibinfo {author} {\bibfnamefont {K.}~\bibnamefont {Lee}}, \bibinfo {author} {\bibfnamefont {R.-Y.}\ \bibnamefont {Zhang}}, \bibinfo {author} {\bibfnamefont {H.-C.}\ \bibnamefont {Park}}, \bibinfo {author} {\bibfnamefont {J.-W.}\ \bibnamefont {Ryu}}, \bibinfo {author} {\bibfnamefont {G.~Y.}\ \bibnamefont {Cho}}, \bibinfo {author} {\bibfnamefont {M.~Y.}\ \bibnamefont {Lee}}, \bibinfo {author} {\bibfnamefont {Z.}~\bibnamefont {Zhang}}, \bibinfo {author} {\bibfnamefont {N.}~\bibnamefont {Park}}, \bibinfo {author} {\bibfnamefont {W.}~\bibnamefont {Jeon}}, \emph {et~al.},\ }\bibfield  {title} {\bibinfo {title} {Spontaneous emission decay and excitation in photonic time crystals},\ }\href@noop {} {\bibfield  {journal} {\bibinfo  {journal} {Physical Review Letters}\ }\textbf {\bibinfo {volume} {135}},\ \bibinfo {pages} {133801} (\bibinfo {year} {2025})}\BibitemShut {NoStop}%
\bibitem [{\citenamefont {Lee}\ \emph {et~al.}(2026)\citenamefont {Lee}, \citenamefont {Kyung}, \citenamefont {Kim}, \citenamefont {Park}, \citenamefont {Lee}, \citenamefont {Choi}, \citenamefont {Chan}, \citenamefont {Shin}, \citenamefont {Kim},\ and\ \citenamefont {Min}}]{lee2026analogs}%
  \BibitemOpen
  \bibfield  {author} {\bibinfo {author} {\bibfnamefont {K.}~\bibnamefont {Lee}}, \bibinfo {author} {\bibfnamefont {M.}~\bibnamefont {Kyung}}, \bibinfo {author} {\bibfnamefont {Y.}~\bibnamefont {Kim}}, \bibinfo {author} {\bibfnamefont {J.}~\bibnamefont {Park}}, \bibinfo {author} {\bibfnamefont {H.}~\bibnamefont {Lee}}, \bibinfo {author} {\bibfnamefont {J.}~\bibnamefont {Choi}}, \bibinfo {author} {\bibfnamefont {C.}~\bibnamefont {Chan}}, \bibinfo {author} {\bibfnamefont {J.}~\bibnamefont {Shin}}, \bibinfo {author} {\bibfnamefont {K.~W.}\ \bibnamefont {Kim}},\ and\ \bibinfo {author} {\bibfnamefont {B.}~\bibnamefont {Min}},\ }\bibfield  {title} {\bibinfo {title} {Analogs of spontaneous emission and lasing in photonic time crystals},\ }\href@noop {} {\bibfield  {journal} {\bibinfo  {journal} {Physical Review Letters}\ }\textbf {\bibinfo {volume} {136}},\ \bibinfo {pages} {093802} (\bibinfo {year} {2026})}\BibitemShut {NoStop}%
\bibitem [{\citenamefont {Huang}\ \emph {et~al.}(2026)\citenamefont {Huang}, \citenamefont {Zhang}, \citenamefont {Zou}, \citenamefont {Bao}, \citenamefont {Di}, \citenamefont {Qin}, \citenamefont {Qian}, \citenamefont {Sun},\ and\ \citenamefont {Zhang}}]{huang2026microwave}%
  \BibitemOpen
  \bibfield  {author} {\bibinfo {author} {\bibfnamefont {L.}~\bibnamefont {Huang}}, \bibinfo {author} {\bibfnamefont {W.}~\bibnamefont {Zhang}}, \bibinfo {author} {\bibfnamefont {D.}~\bibnamefont {Zou}}, \bibinfo {author} {\bibfnamefont {J.}~\bibnamefont {Bao}}, \bibinfo {author} {\bibfnamefont {F.}~\bibnamefont {Di}}, \bibinfo {author} {\bibfnamefont {H.}~\bibnamefont {Qin}}, \bibinfo {author} {\bibfnamefont {L.}~\bibnamefont {Qian}}, \bibinfo {author} {\bibfnamefont {H.}~\bibnamefont {Sun}},\ and\ \bibinfo {author} {\bibfnamefont {X.}~\bibnamefont {Zhang}},\ }\bibfield  {title} {\bibinfo {title} {Microwave vortex beam lasing via photonic time crystals},\ }\href@noop {} {\bibfield  {journal} {\bibinfo  {journal} {Physical Review Letters}\ }\textbf {\bibinfo {volume} {137}},\ \bibinfo {pages} {023801} (\bibinfo {year} {2026})}\BibitemShut {NoStop}%
\bibitem [{\citenamefont {Lustig}\ \emph {et~al.}(2018)\citenamefont {Lustig}, \citenamefont {Sharabi},\ and\ \citenamefont {Segev}}]{lustig2018topological}%
  \BibitemOpen
  \bibfield  {author} {\bibinfo {author} {\bibfnamefont {E.}~\bibnamefont {Lustig}}, \bibinfo {author} {\bibfnamefont {Y.}~\bibnamefont {Sharabi}},\ and\ \bibinfo {author} {\bibfnamefont {M.}~\bibnamefont {Segev}},\ }\bibfield  {title} {\bibinfo {title} {Topological aspects of photonic time crystals},\ }\href@noop {} {\bibfield  {journal} {\bibinfo  {journal} {Optica}\ }\textbf {\bibinfo {volume} {5}},\ \bibinfo {pages} {1390} (\bibinfo {year} {2018})}\BibitemShut {NoStop}%
\bibitem [{\citenamefont {Ma}\ and\ \citenamefont {Wang}(2019)}]{ma2019band}%
  \BibitemOpen
  \bibfield  {author} {\bibinfo {author} {\bibfnamefont {J.}~\bibnamefont {Ma}}\ and\ \bibinfo {author} {\bibfnamefont {Z.-G.}\ \bibnamefont {Wang}},\ }\bibfield  {title} {\bibinfo {title} {Band structure and topological phase transition of photonic time crystals},\ }\href@noop {} {\bibfield  {journal} {\bibinfo  {journal} {Optics Express}\ }\textbf {\bibinfo {volume} {27}},\ \bibinfo {pages} {12914} (\bibinfo {year} {2019})}\BibitemShut {NoStop}%
\bibitem [{\citenamefont {Dong}\ \emph {et~al.}(2023)\citenamefont {Dong}, \citenamefont {Liu}, \citenamefont {Sui},\ and\ \citenamefont {Zhang}}]{dong2023band}%
  \BibitemOpen
  \bibfield  {author} {\bibinfo {author} {\bibfnamefont {R.-Y.}\ \bibnamefont {Dong}}, \bibinfo {author} {\bibfnamefont {Y.-M.}\ \bibnamefont {Liu}}, \bibinfo {author} {\bibfnamefont {J.-Y.}\ \bibnamefont {Sui}},\ and\ \bibinfo {author} {\bibfnamefont {H.-F.}\ \bibnamefont {Zhang}},\ }\bibfield  {title} {\bibinfo {title} {Band structure and temporal topological edge state of continuous photonic time crystals},\ }\href@noop {} {\bibfield  {journal} {\bibinfo  {journal} {IEEE Transactions on Antennas and Propagation}\ }\textbf {\bibinfo {volume} {72}},\ \bibinfo {pages} {674} (\bibinfo {year} {2023})}\BibitemShut {NoStop}%
\bibitem [{\citenamefont {He}\ \emph {et~al.}(2026)\citenamefont {He}, \citenamefont {Zhang}, \citenamefont {Li},\ and\ \citenamefont {Ni}}]{he2026temporal}%
  \BibitemOpen
  \bibfield  {author} {\bibinfo {author} {\bibfnamefont {Z.}~\bibnamefont {He}}, \bibinfo {author} {\bibfnamefont {S.}~\bibnamefont {Zhang}}, \bibinfo {author} {\bibfnamefont {H.}~\bibnamefont {Li}},\ and\ \bibinfo {author} {\bibfnamefont {X.}~\bibnamefont {Ni}},\ }\bibfield  {title} {\bibinfo {title} {Temporal weyl physics and topological control of direction-selected radiation in anisotropic photonic time crystals},\ }\href@noop {} {\bibfield  {journal} {\bibinfo  {journal} {Physical Review B}\ }\textbf {\bibinfo {volume} {113}},\ \bibinfo {pages} {205129} (\bibinfo {year} {2026})}\BibitemShut {NoStop}%
\bibitem [{\citenamefont {Dikopoltsev}\ \emph {et~al.}(2022)\citenamefont {Dikopoltsev}, \citenamefont {Sharabi}, \citenamefont {Lyubarov}, \citenamefont {Lumer}, \citenamefont {Tsesses}, \citenamefont {Lustig}, \citenamefont {Kaminer},\ and\ \citenamefont {Segev}}]{dikopoltsev2022light}%
  \BibitemOpen
  \bibfield  {author} {\bibinfo {author} {\bibfnamefont {A.}~\bibnamefont {Dikopoltsev}}, \bibinfo {author} {\bibfnamefont {Y.}~\bibnamefont {Sharabi}}, \bibinfo {author} {\bibfnamefont {M.}~\bibnamefont {Lyubarov}}, \bibinfo {author} {\bibfnamefont {Y.}~\bibnamefont {Lumer}}, \bibinfo {author} {\bibfnamefont {S.}~\bibnamefont {Tsesses}}, \bibinfo {author} {\bibfnamefont {E.}~\bibnamefont {Lustig}}, \bibinfo {author} {\bibfnamefont {I.}~\bibnamefont {Kaminer}},\ and\ \bibinfo {author} {\bibfnamefont {M.}~\bibnamefont {Segev}},\ }\bibfield  {title} {\bibinfo {title} {Light emission by free electrons in photonic time-crystals},\ }\href@noop {} {\bibfield  {journal} {\bibinfo  {journal} {Proceedings of the National Academy of Sciences}\ }\textbf {\bibinfo {volume} {119}},\ \bibinfo {pages} {e2119705119} (\bibinfo {year} {2022})}\BibitemShut {NoStop}%
\bibitem [{\citenamefont {Gao}\ \emph {et~al.}(2024)\citenamefont {Gao}, \citenamefont {Zhao}, \citenamefont {Ma},\ and\ \citenamefont {Dong}}]{gao2024free}%
  \BibitemOpen
  \bibfield  {author} {\bibinfo {author} {\bibfnamefont {X.}~\bibnamefont {Gao}}, \bibinfo {author} {\bibfnamefont {X.}~\bibnamefont {Zhao}}, \bibinfo {author} {\bibfnamefont {X.}~\bibnamefont {Ma}},\ and\ \bibinfo {author} {\bibfnamefont {T.}~\bibnamefont {Dong}},\ }\bibfield  {title} {\bibinfo {title} {Free electron emission in vacuum assisted by photonic time crystals},\ }\href@noop {} {\bibfield  {journal} {\bibinfo  {journal} {Journal of Physics D: Applied Physics}\ }\textbf {\bibinfo {volume} {57}},\ \bibinfo {pages} {315112} (\bibinfo {year} {2024})}\BibitemShut {NoStop}%
\bibitem [{\citenamefont {Yang}\ \emph {et~al.}(2025)\citenamefont {Yang}, \citenamefont {Hu}, \citenamefont {Liu}, \citenamefont {Yang}, \citenamefont {Yu}, \citenamefont {Long}, \citenamefont {Zheng}, \citenamefont {Luo}, \citenamefont {Li},\ and\ \citenamefont {Garcia-Vidal}}]{yang2025topologically}%
  \BibitemOpen
  \bibfield  {author} {\bibinfo {author} {\bibfnamefont {Y.}~\bibnamefont {Yang}}, \bibinfo {author} {\bibfnamefont {H.}~\bibnamefont {Hu}}, \bibinfo {author} {\bibfnamefont {L.}~\bibnamefont {Liu}}, \bibinfo {author} {\bibfnamefont {Y.}~\bibnamefont {Yang}}, \bibinfo {author} {\bibfnamefont {Y.}~\bibnamefont {Yu}}, \bibinfo {author} {\bibfnamefont {Y.}~\bibnamefont {Long}}, \bibinfo {author} {\bibfnamefont {X.}~\bibnamefont {Zheng}}, \bibinfo {author} {\bibfnamefont {Y.}~\bibnamefont {Luo}}, \bibinfo {author} {\bibfnamefont {Z.}~\bibnamefont {Li}},\ and\ \bibinfo {author} {\bibfnamefont {F.~J.}\ \bibnamefont {Garcia-Vidal}},\ }\bibfield  {title} {\bibinfo {title} {Topologically protected edge states in time photonic crystals with chiral symmetry},\ }\href@noop {} {\bibfield  {journal} {\bibinfo  {journal} {Acs Photonics}\ }\textbf {\bibinfo {volume} {12}},\ \bibinfo {pages} {2389} (\bibinfo {year} {2025})}\BibitemShut {NoStop}%
\bibitem [{\citenamefont {Xiong}\ \emph {et~al.}(2025)\citenamefont {Xiong}, \citenamefont {Zhang}, \citenamefont {Duan}, \citenamefont {Wang}, \citenamefont {Long}, \citenamefont {Hou}, \citenamefont {Yu}, \citenamefont {Zou},\ and\ \citenamefont {Zhang}}]{xiong2025observation}%
  \BibitemOpen
  \bibfield  {author} {\bibinfo {author} {\bibfnamefont {J.}~\bibnamefont {Xiong}}, \bibinfo {author} {\bibfnamefont {X.}~\bibnamefont {Zhang}}, \bibinfo {author} {\bibfnamefont {L.}~\bibnamefont {Duan}}, \bibinfo {author} {\bibfnamefont {J.}~\bibnamefont {Wang}}, \bibinfo {author} {\bibfnamefont {Y.}~\bibnamefont {Long}}, \bibinfo {author} {\bibfnamefont {H.}~\bibnamefont {Hou}}, \bibinfo {author} {\bibfnamefont {L.}~\bibnamefont {Yu}}, \bibinfo {author} {\bibfnamefont {L.}~\bibnamefont {Zou}},\ and\ \bibinfo {author} {\bibfnamefont {B.}~\bibnamefont {Zhang}},\ }\bibfield  {title} {\bibinfo {title} {Observation of wave amplification and temporal topological state in a non-synthetic photonic time crystal},\ }\href@noop {} {\bibfield  {journal} {\bibinfo  {journal} {Nature Communications}\ }\textbf {\bibinfo {volume} {16}},\ \bibinfo {pages} {11182} (\bibinfo {year} {2025})}\BibitemShut {NoStop}%
\bibitem [{\citenamefont {Jiang}\ \emph {et~al.}(2026)\citenamefont {Jiang}, \citenamefont {Hu}, \citenamefont {Long}, \citenamefont {Liu}, \citenamefont {Hou}, \citenamefont {Liu},\ and\ \citenamefont {Li}}]{jiang2026broadband}%
  \BibitemOpen
  \bibfield  {author} {\bibinfo {author} {\bibfnamefont {J.}~\bibnamefont {Jiang}}, \bibinfo {author} {\bibfnamefont {H.}~\bibnamefont {Hu}}, \bibinfo {author} {\bibfnamefont {Y.}~\bibnamefont {Long}}, \bibinfo {author} {\bibfnamefont {L.}~\bibnamefont {Liu}}, \bibinfo {author} {\bibfnamefont {S.}~\bibnamefont {Hou}}, \bibinfo {author} {\bibfnamefont {D.}~\bibnamefont {Liu}},\ and\ \bibinfo {author} {\bibfnamefont {Z.}~\bibnamefont {Li}},\ }\bibfield  {title} {\bibinfo {title} {Broadband temporal localization and delocalized temporal edge states in time photonic crystals},\ }\href@noop {} {\bibfield  {journal} {\bibinfo  {journal} {Physical Review A}\ }\textbf {\bibinfo {volume} {113}},\ \bibinfo {pages} {043520} (\bibinfo {year} {2026})}\BibitemShut {NoStop}%
\bibitem [{\citenamefont {Wang}\ \emph {et~al.}(2018)\citenamefont {Wang}, \citenamefont {Zhang},\ and\ \citenamefont {Chan}}]{wang2018photonic}%
  \BibitemOpen
  \bibfield  {author} {\bibinfo {author} {\bibfnamefont {N.}~\bibnamefont {Wang}}, \bibinfo {author} {\bibfnamefont {Z.-Q.}\ \bibnamefont {Zhang}},\ and\ \bibinfo {author} {\bibfnamefont {C.~T.}\ \bibnamefont {Chan}},\ }\bibfield  {title} {\bibinfo {title} {Photonic floquet media with a complex time-periodic permittivity},\ }\href@noop {} {\bibfield  {journal} {\bibinfo  {journal} {Physical Review B}\ }\textbf {\bibinfo {volume} {98}},\ \bibinfo {pages} {085142} (\bibinfo {year} {2018})}\BibitemShut {NoStop}%
\bibitem [{\citenamefont {Hayran}\ and\ \citenamefont {Monticone}(2021)}]{hayran2021controlling}%
  \BibitemOpen
  \bibfield  {author} {\bibinfo {author} {\bibfnamefont {Z.}~\bibnamefont {Hayran}}\ and\ \bibinfo {author} {\bibfnamefont {F.}~\bibnamefont {Monticone}},\ }\bibfield  {title} {\bibinfo {title} {Controlling the spectral flow of light in non-hermitian photonic time crystals},\ }in\ \href@noop {} {\emph {\bibinfo {booktitle} {2021 Fifteenth International Congress on Artificial Materials for Novel Wave Phenomena (Metamaterials)}}}\ (\bibinfo {organization} {IEEE},\ \bibinfo {year} {2021})\ pp.\ \bibinfo {pages} {153--155}\BibitemShut {NoStop}%
\bibitem [{\citenamefont {Zhang}\ \emph {et~al.}(2025)\citenamefont {Zhang}, \citenamefont {Yang}, \citenamefont {Sha}, \citenamefont {Xie},\ and\ \citenamefont {Yang}}]{zhang2025parity}%
  \BibitemOpen
  \bibfield  {author} {\bibinfo {author} {\bibfnamefont {R.-C.}\ \bibnamefont {Zhang}}, \bibinfo {author} {\bibfnamefont {S.}~\bibnamefont {Yang}}, \bibinfo {author} {\bibfnamefont {Y.}~\bibnamefont {Sha}}, \bibinfo {author} {\bibfnamefont {Z.}~\bibnamefont {Xie}},\ and\ \bibinfo {author} {\bibfnamefont {Y.}~\bibnamefont {Yang}},\ }\bibfield  {title} {\bibinfo {title} {Parity-time symmetry phase transition in photonic time-modulated media},\ }\href@noop {} {\bibfield  {journal} {\bibinfo  {journal} {arXiv preprint arXiv:2507.03337}\ } (\bibinfo {year} {2025})}\BibitemShut {NoStop}%
\bibitem [{\citenamefont {Zhou}\ \emph {et~al.}(2025)\citenamefont {Zhou}, \citenamefont {Yan}, \citenamefont {Lu},\ and\ \citenamefont {Hu}}]{zhou2025band}%
  \BibitemOpen
  \bibfield  {author} {\bibinfo {author} {\bibfnamefont {R.}~\bibnamefont {Zhou}}, \bibinfo {author} {\bibfnamefont {Q.}~\bibnamefont {Yan}}, \bibinfo {author} {\bibfnamefont {C.}~\bibnamefont {Lu}},\ and\ \bibinfo {author} {\bibfnamefont {X.}~\bibnamefont {Hu}},\ }\bibfield  {title} {\bibinfo {title} {Band structures of lossy photonic time crystals},\ }in\ \href@noop {} {\emph {\bibinfo {booktitle} {Fifth Optics Frontier Conference (OFS 2025)}}},\ Vol.\ \bibinfo {volume} {13648}\ (\bibinfo {organization} {SPIE},\ \bibinfo {year} {2025})\ pp.\ \bibinfo {pages} {57--62}\BibitemShut {NoStop}%
\bibitem [{\citenamefont {Galiffi}\ \emph {et~al.}(2025)\citenamefont {Galiffi}, \citenamefont {Solís}, \citenamefont {Yin}, \citenamefont {Engheta},\ and\ \citenamefont {Alù}}]{galiffi2025electrodynamics}%
  \BibitemOpen
  \bibfield  {author} {\bibinfo {author} {\bibfnamefont {E.}~\bibnamefont {Galiffi}}, \bibinfo {author} {\bibfnamefont {D.~M.}\ \bibnamefont {Solís}}, \bibinfo {author} {\bibfnamefont {S.}~\bibnamefont {Yin}}, \bibinfo {author} {\bibfnamefont {N.}~\bibnamefont {Engheta}},\ and\ \bibinfo {author} {\bibfnamefont {A.}~\bibnamefont {Alù}},\ }\bibfield  {title} {\bibinfo {title} {Electrodynamics of photonic temporal interfaces},\ }\href@noop {} {\bibfield  {journal} {\bibinfo  {journal} {Light: Science \& Applications}\ }\textbf {\bibinfo {volume} {14}},\ \bibinfo {pages} {338} (\bibinfo {year} {2025})}\BibitemShut {NoStop}%
\bibitem [{\citenamefont {Harfoush}\ and\ \citenamefont {Taflove}(1991)}]{harfoush1991scattering}%
  \BibitemOpen
  \bibfield  {author} {\bibinfo {author} {\bibfnamefont {F.}~\bibnamefont {Harfoush}}\ and\ \bibinfo {author} {\bibfnamefont {A.}~\bibnamefont {Taflove}},\ }\bibfield  {title} {\bibinfo {title} {Scattering of electromagnetic waves by a material half-space with a time-varying conductivity},\ }\href@noop {} {\bibfield  {journal} {\bibinfo  {journal} {IEEE transactions on antennas and propagation}\ }\textbf {\bibinfo {volume} {39}},\ \bibinfo {pages} {898} (\bibinfo {year} {1991})}\BibitemShut {NoStop}%
\bibitem [{\citenamefont {Dinc}\ \emph {et~al.}(2017)\citenamefont {Dinc}, \citenamefont {Tymchenko}, \citenamefont {Nagulu}, \citenamefont {Sounas}, \citenamefont {Alu},\ and\ \citenamefont {Krishnaswamy}}]{dinc2017synchronized}%
  \BibitemOpen
  \bibfield  {author} {\bibinfo {author} {\bibfnamefont {T.}~\bibnamefont {Dinc}}, \bibinfo {author} {\bibfnamefont {M.}~\bibnamefont {Tymchenko}}, \bibinfo {author} {\bibfnamefont {A.}~\bibnamefont {Nagulu}}, \bibinfo {author} {\bibfnamefont {D.}~\bibnamefont {Sounas}}, \bibinfo {author} {\bibfnamefont {A.}~\bibnamefont {Alu}},\ and\ \bibinfo {author} {\bibfnamefont {H.}~\bibnamefont {Krishnaswamy}},\ }\bibfield  {title} {\bibinfo {title} {Synchronized conductivity modulation to realize broadband lossless magnetic-free non-reciprocity},\ }\href@noop {} {\bibfield  {journal} {\bibinfo  {journal} {Nature communications}\ }\textbf {\bibinfo {volume} {8}},\ \bibinfo {pages} {795} (\bibinfo {year} {2017})}\BibitemShut {NoStop}%
\bibitem [{\citenamefont {Song}\ \emph {et~al.}(2019)\citenamefont {Song}, \citenamefont {Shi}, \citenamefont {Lin},\ and\ \citenamefont {Fan}}]{song2019direction}%
  \BibitemOpen
  \bibfield  {author} {\bibinfo {author} {\bibfnamefont {A.~Y.}\ \bibnamefont {Song}}, \bibinfo {author} {\bibfnamefont {Y.}~\bibnamefont {Shi}}, \bibinfo {author} {\bibfnamefont {Q.}~\bibnamefont {Lin}},\ and\ \bibinfo {author} {\bibfnamefont {S.}~\bibnamefont {Fan}},\ }\bibfield  {title} {\bibinfo {title} {Direction-dependent parity-time phase transition and nonreciprocal amplification with dynamic gain-loss modulation},\ }\href@noop {} {\bibfield  {journal} {\bibinfo  {journal} {Physical Review A}\ }\textbf {\bibinfo {volume} {99}},\ \bibinfo {pages} {013824} (\bibinfo {year} {2019})}\BibitemShut {NoStop}%
\bibitem [{\citenamefont {Li}\ \emph {et~al.}(2021)\citenamefont {Li}, \citenamefont {Yin}, \citenamefont {Galiffi},\ and\ \citenamefont {Al{\`u}}}]{li2021temporal}%
  \BibitemOpen
  \bibfield  {author} {\bibinfo {author} {\bibfnamefont {H.}~\bibnamefont {Li}}, \bibinfo {author} {\bibfnamefont {S.}~\bibnamefont {Yin}}, \bibinfo {author} {\bibfnamefont {E.}~\bibnamefont {Galiffi}},\ and\ \bibinfo {author} {\bibfnamefont {A.}~\bibnamefont {Al{\`u}}},\ }\bibfield  {title} {\bibinfo {title} {Temporal parity-time symmetry for extreme energy transformations},\ }\href@noop {} {\bibfield  {journal} {\bibinfo  {journal} {Physical Review Letters}\ }\textbf {\bibinfo {volume} {127}},\ \bibinfo {pages} {153903} (\bibinfo {year} {2021})}\BibitemShut {NoStop}%
\bibitem [{sup()}]{suppmat}%
  \BibitemOpen
  \href@noop {} {}\bibinfo {note} {See Supplemental Material for the deviation of Floquet dispersion relation, selection of momentum for topological edge states, variation of eigenstate with momentum in the Bloch sphere, temporal evolution of field amplitude around critical angle, verification of topological protection and derivation of Dirac mass.}\BibitemShut {Stop}%
\bibitem [{\citenamefont {Tong}\ \emph {et~al.}(2025)\citenamefont {Tong}, \citenamefont {Zhang}, \citenamefont {Li}, \citenamefont {Zhang}, \citenamefont {Xie},\ and\ \citenamefont {Qiu}}]{tong2025observation}%
  \BibitemOpen
  \bibfield  {author} {\bibinfo {author} {\bibfnamefont {S.}~\bibnamefont {Tong}}, \bibinfo {author} {\bibfnamefont {Q.}~\bibnamefont {Zhang}}, \bibinfo {author} {\bibfnamefont {G.}~\bibnamefont {Li}}, \bibinfo {author} {\bibfnamefont {K.}~\bibnamefont {Zhang}}, \bibinfo {author} {\bibfnamefont {C.}~\bibnamefont {Xie}},\ and\ \bibinfo {author} {\bibfnamefont {C.}~\bibnamefont {Qiu}},\ }\bibfield  {title} {\bibinfo {title} {Observation of momentum-band topology in pt-symmetric floquet lattices},\ }\href@noop {} {\bibfield  {journal} {\bibinfo  {journal} {Nature Communications}\ }\textbf {\bibinfo {volume} {16}},\ \bibinfo {pages} {9975} (\bibinfo {year} {2025})}\BibitemShut {NoStop}%
\bibitem [{\citenamefont {Pan}\ \emph {et~al.}(2023)\citenamefont {Pan}, \citenamefont {Cohen},\ and\ \citenamefont {Segev}}]{pan2023superluminal}%
  \BibitemOpen
  \bibfield  {author} {\bibinfo {author} {\bibfnamefont {Y.}~\bibnamefont {Pan}}, \bibinfo {author} {\bibfnamefont {M.-I.}\ \bibnamefont {Cohen}},\ and\ \bibinfo {author} {\bibfnamefont {M.}~\bibnamefont {Segev}},\ }\bibfield  {title} {\bibinfo {title} {Superluminal k-gap solitons in nonlinear photonic time crystals},\ }\href@noop {} {\bibfield  {journal} {\bibinfo  {journal} {Physical Review Letters}\ }\textbf {\bibinfo {volume} {130}},\ \bibinfo {pages} {233801} (\bibinfo {year} {2023})}\BibitemShut {NoStop}%
\bibitem [{\citenamefont {Lu}\ \emph {et~al.}(2018)\citenamefont {Lu}, \citenamefont {Gao},\ and\ \citenamefont {Wang}}]{lu2018topological}%
  \BibitemOpen
  \bibfield  {author} {\bibinfo {author} {\bibfnamefont {L.}~\bibnamefont {Lu}}, \bibinfo {author} {\bibfnamefont {H.}~\bibnamefont {Gao}},\ and\ \bibinfo {author} {\bibfnamefont {Z.}~\bibnamefont {Wang}},\ }\bibfield  {title} {\bibinfo {title} {Topological one-way fiber of second chern number},\ }\href@noop {} {\bibfield  {journal} {\bibinfo  {journal} {Nature communications}\ }\textbf {\bibinfo {volume} {9}},\ \bibinfo {pages} {5384} (\bibinfo {year} {2018})}\BibitemShut {NoStop}%
\bibitem [{\citenamefont {Gao}\ \emph {et~al.}(2020)\citenamefont {Gao}, \citenamefont {Yang}, \citenamefont {Lin}, \citenamefont {Zhang}, \citenamefont {Li}, \citenamefont {Bo}, \citenamefont {Wang},\ and\ \citenamefont {Lu}}]{gao2020dirac}%
  \BibitemOpen
  \bibfield  {author} {\bibinfo {author} {\bibfnamefont {X.}~\bibnamefont {Gao}}, \bibinfo {author} {\bibfnamefont {L.}~\bibnamefont {Yang}}, \bibinfo {author} {\bibfnamefont {H.}~\bibnamefont {Lin}}, \bibinfo {author} {\bibfnamefont {L.}~\bibnamefont {Zhang}}, \bibinfo {author} {\bibfnamefont {J.}~\bibnamefont {Li}}, \bibinfo {author} {\bibfnamefont {F.}~\bibnamefont {Bo}}, \bibinfo {author} {\bibfnamefont {Z.}~\bibnamefont {Wang}},\ and\ \bibinfo {author} {\bibfnamefont {L.}~\bibnamefont {Lu}},\ }\bibfield  {title} {\bibinfo {title} {Dirac-vortex topological cavities},\ }\href@noop {} {\bibfield  {journal} {\bibinfo  {journal} {Nature Nanotechnology}\ }\textbf {\bibinfo {volume} {15}},\ \bibinfo {pages} {1012} (\bibinfo {year} {2020})}\BibitemShut {NoStop}%
\bibitem [{\citenamefont {Bi}\ and\ \citenamefont {Wang}(2015)}]{bi2015unidirectional}%
  \BibitemOpen
  \bibfield  {author} {\bibinfo {author} {\bibfnamefont {R.}~\bibnamefont {Bi}}\ and\ \bibinfo {author} {\bibfnamefont {Z.}~\bibnamefont {Wang}},\ }\bibfield  {title} {\bibinfo {title} {Unidirectional transport in electronic and photonic weyl materials by dirac mass engineering},\ }\href@noop {} {\bibfield  {journal} {\bibinfo  {journal} {Physical Review B}\ }\textbf {\bibinfo {volume} {92}},\ \bibinfo {pages} {241109} (\bibinfo {year} {2015})}\BibitemShut {NoStop}%
\bibitem [{\citenamefont {Ren}\ \emph {et~al.}(2025)\citenamefont {Ren}, \citenamefont {Ye}, \citenamefont {Chen}, \citenamefont {Chen}, \citenamefont {Zhang}, \citenamefont {Pan}, \citenamefont {Li}, \citenamefont {Li}, \citenamefont {Zhang}, \citenamefont {Chen} \emph {et~al.}}]{ren2025observation}%
  \BibitemOpen
  \bibfield  {author} {\bibinfo {author} {\bibfnamefont {Y.}~\bibnamefont {Ren}}, \bibinfo {author} {\bibfnamefont {K.}~\bibnamefont {Ye}}, \bibinfo {author} {\bibfnamefont {Q.}~\bibnamefont {Chen}}, \bibinfo {author} {\bibfnamefont {F.}~\bibnamefont {Chen}}, \bibinfo {author} {\bibfnamefont {L.}~\bibnamefont {Zhang}}, \bibinfo {author} {\bibfnamefont {Y.}~\bibnamefont {Pan}}, \bibinfo {author} {\bibfnamefont {W.}~\bibnamefont {Li}}, \bibinfo {author} {\bibfnamefont {X.}~\bibnamefont {Li}}, \bibinfo {author} {\bibfnamefont {L.}~\bibnamefont {Zhang}}, \bibinfo {author} {\bibfnamefont {H.}~\bibnamefont {Chen}}, \emph {et~al.},\ }\bibfield  {title} {\bibinfo {title} {Observation of momentum-gap topology of light at temporal interfaces in a time-synthetic lattice},\ }\href@noop {} {\bibfield  {journal} {\bibinfo  {journal} {Nature Communications}\ }\textbf {\bibinfo {volume} {16}},\ \bibinfo {pages} {707} (\bibinfo {year} {2025})}\BibitemShut {NoStop}%
\bibitem [{\citenamefont {Li}\ \emph {et~al.}(2025)\citenamefont {Li}, \citenamefont {Liu}, \citenamefont {Wang}, \citenamefont {Chen}, \citenamefont {Ma},\ and\ \citenamefont {Dong}}]{li2025topological}%
  \BibitemOpen
  \bibfield  {author} {\bibinfo {author} {\bibfnamefont {M.-W.}\ \bibnamefont {Li}}, \bibinfo {author} {\bibfnamefont {J.-W.}\ \bibnamefont {Liu}}, \bibinfo {author} {\bibfnamefont {X.}~\bibnamefont {Wang}}, \bibinfo {author} {\bibfnamefont {W.-J.}\ \bibnamefont {Chen}}, \bibinfo {author} {\bibfnamefont {G.}~\bibnamefont {Ma}},\ and\ \bibinfo {author} {\bibfnamefont {J.-W.}\ \bibnamefont {Dong}},\ }\bibfield  {title} {\bibinfo {title} {Topological temporal boundary states in a non-hermitian spatial crystal},\ }\href@noop {} {\bibfield  {journal} {\bibinfo  {journal} {Physical Review Letters}\ }\textbf {\bibinfo {volume} {135}},\ \bibinfo {pages} {187101} (\bibinfo {year} {2025})}\BibitemShut {NoStop}%
\bibitem [{\citenamefont {Sharabi}\ \emph {et~al.}(2022)\citenamefont {Sharabi}, \citenamefont {Dikopoltsev}, \citenamefont {Lustig}, \citenamefont {Lumer},\ and\ \citenamefont {Segev}}]{sharabi2022spatiotemporal}%
  \BibitemOpen
  \bibfield  {author} {\bibinfo {author} {\bibfnamefont {Y.}~\bibnamefont {Sharabi}}, \bibinfo {author} {\bibfnamefont {A.}~\bibnamefont {Dikopoltsev}}, \bibinfo {author} {\bibfnamefont {E.}~\bibnamefont {Lustig}}, \bibinfo {author} {\bibfnamefont {Y.}~\bibnamefont {Lumer}},\ and\ \bibinfo {author} {\bibfnamefont {M.}~\bibnamefont {Segev}},\ }\bibfield  {title} {\bibinfo {title} {Spatiotemporal photonic crystals},\ }\href@noop {} {\bibfield  {journal} {\bibinfo  {journal} {Optica}\ }\textbf {\bibinfo {volume} {9}},\ \bibinfo {pages} {585} (\bibinfo {year} {2022})}\BibitemShut {NoStop}%
\bibitem [{\citenamefont {Segal}\ \emph {et~al.}(2025)\citenamefont {Segal}, \citenamefont {Plotnik}, \citenamefont {Lustig}, \citenamefont {Sharabi}, \citenamefont {Cohen}, \citenamefont {Dikopoltsev},\ and\ \citenamefont {Segev}}]{segal2025two}%
  \BibitemOpen
  \bibfield  {author} {\bibinfo {author} {\bibfnamefont {O.}~\bibnamefont {Segal}}, \bibinfo {author} {\bibfnamefont {Y.}~\bibnamefont {Plotnik}}, \bibinfo {author} {\bibfnamefont {E.}~\bibnamefont {Lustig}}, \bibinfo {author} {\bibfnamefont {Y.}~\bibnamefont {Sharabi}}, \bibinfo {author} {\bibfnamefont {M.-I.}\ \bibnamefont {Cohen}}, \bibinfo {author} {\bibfnamefont {A.}~\bibnamefont {Dikopoltsev}},\ and\ \bibinfo {author} {\bibfnamefont {M.}~\bibnamefont {Segev}},\ }\bibfield  {title} {\bibinfo {title} {Two-dimensional topological edge states in periodic space-time interfaces},\ }\href@noop {} {\bibfield  {journal} {\bibinfo  {journal} {Physical Review Letters}\ }\textbf {\bibinfo {volume} {135}},\ \bibinfo {pages} {163801} (\bibinfo {year} {2025})}\BibitemShut {NoStop}%
\bibitem [{\citenamefont {Zhang}\ \emph {et~al.}(2026)\citenamefont {Zhang}, \citenamefont {Zhao}, \citenamefont {Pan}, \citenamefont {Pan}, \citenamefont {Cheng},\ and\ \citenamefont {Pan}}]{zhang2026topological}%
  \BibitemOpen
  \bibfield  {author} {\bibinfo {author} {\bibfnamefont {L.}~\bibnamefont {Zhang}}, \bibinfo {author} {\bibfnamefont {Z.}~\bibnamefont {Zhao}}, \bibinfo {author} {\bibfnamefont {Q.}~\bibnamefont {Pan}}, \bibinfo {author} {\bibfnamefont {C.}~\bibnamefont {Pan}}, \bibinfo {author} {\bibfnamefont {Q.}~\bibnamefont {Cheng}},\ and\ \bibinfo {author} {\bibfnamefont {Y.}~\bibnamefont {Pan}},\ }\bibfield  {title} {\bibinfo {title} {Topological event wavepackets in energy-momentum-gapped photonic spacetime crystals},\ }\href@noop {} {\bibfield  {journal} {\bibinfo  {journal} {Laser \& Photonics Reviews}\ ,\ \bibinfo {pages} {e02603}} (\bibinfo {year} {2026})}\BibitemShut {NoStop}%
\bibitem [{\citenamefont {Dong}\ \emph {et~al.}(2024)\citenamefont {Dong}, \citenamefont {Li}, \citenamefont {Wan}, \citenamefont {Liang}, \citenamefont {Yang},\ and\ \citenamefont {Yan}}]{dong2024quantum}%
  \BibitemOpen
  \bibfield  {author} {\bibinfo {author} {\bibfnamefont {Z.}~\bibnamefont {Dong}}, \bibinfo {author} {\bibfnamefont {H.}~\bibnamefont {Li}}, \bibinfo {author} {\bibfnamefont {T.}~\bibnamefont {Wan}}, \bibinfo {author} {\bibfnamefont {Q.}~\bibnamefont {Liang}}, \bibinfo {author} {\bibfnamefont {Z.}~\bibnamefont {Yang}},\ and\ \bibinfo {author} {\bibfnamefont {B.}~\bibnamefont {Yan}},\ }\bibfield  {title} {\bibinfo {title} {Quantum time reflection and refraction of ultracold atoms},\ }\href@noop {} {\bibfield  {journal} {\bibinfo  {journal} {Nature Photonics}\ }\textbf {\bibinfo {volume} {18}},\ \bibinfo {pages} {68} (\bibinfo {year} {2024})}\BibitemShut {NoStop}%
\bibitem [{\citenamefont {Yu}\ \emph {et~al.}(2025)\citenamefont {Yu}, \citenamefont {Song}, \citenamefont {Wang}, \citenamefont {Srikanth}, \citenamefont {Kaushik~Sridhar}, \citenamefont {Chen}, \citenamefont {Huang}, \citenamefont {Li}, \citenamefont {Qiao}, \citenamefont {Wu} \emph {et~al.}}]{yu2025comprehensive}%
  \BibitemOpen
  \bibfield  {author} {\bibinfo {author} {\bibfnamefont {D.}~\bibnamefont {Yu}}, \bibinfo {author} {\bibfnamefont {W.}~\bibnamefont {Song}}, \bibinfo {author} {\bibfnamefont {L.}~\bibnamefont {Wang}}, \bibinfo {author} {\bibfnamefont {R.}~\bibnamefont {Srikanth}}, \bibinfo {author} {\bibfnamefont {S.}~\bibnamefont {Kaushik~Sridhar}}, \bibinfo {author} {\bibfnamefont {T.}~\bibnamefont {Chen}}, \bibinfo {author} {\bibfnamefont {C.}~\bibnamefont {Huang}}, \bibinfo {author} {\bibfnamefont {G.}~\bibnamefont {Li}}, \bibinfo {author} {\bibfnamefont {X.}~\bibnamefont {Qiao}}, \bibinfo {author} {\bibfnamefont {X.}~\bibnamefont {Wu}}, \emph {et~al.},\ }\bibfield  {title} {\bibinfo {title} {Comprehensive review on developments of synthetic dimensions},\ }\href@noop {} {\bibfield  {journal} {\bibinfo  {journal} {Photonics Insights}\ }\textbf {\bibinfo {volume} {4}},\ \bibinfo {pages} {R06} (\bibinfo {year} {2025})}\BibitemShut {NoStop}%
\bibitem [{\citenamefont {Ye}\ \emph {et~al.}(2023)\citenamefont {Ye}, \citenamefont {Qin}, \citenamefont {Wang}, \citenamefont {Zhao}, \citenamefont {Liu}, \citenamefont {Wang}, \citenamefont {Longhi},\ and\ \citenamefont {Lu}}]{ye2023reconfigurable}%
  \BibitemOpen
  \bibfield  {author} {\bibinfo {author} {\bibfnamefont {H.}~\bibnamefont {Ye}}, \bibinfo {author} {\bibfnamefont {C.}~\bibnamefont {Qin}}, \bibinfo {author} {\bibfnamefont {S.}~\bibnamefont {Wang}}, \bibinfo {author} {\bibfnamefont {L.}~\bibnamefont {Zhao}}, \bibinfo {author} {\bibfnamefont {W.}~\bibnamefont {Liu}}, \bibinfo {author} {\bibfnamefont {B.}~\bibnamefont {Wang}}, \bibinfo {author} {\bibfnamefont {S.}~\bibnamefont {Longhi}},\ and\ \bibinfo {author} {\bibfnamefont {P.}~\bibnamefont {Lu}},\ }\bibfield  {title} {\bibinfo {title} {Reconfigurable refraction manipulation at synthetic temporal interfaces with scalar and vector gauge potentials},\ }\href@noop {} {\bibfield  {journal} {\bibinfo  {journal} {Proceedings of the National Academy of Sciences}\ }\textbf {\bibinfo {volume} {120}},\ \bibinfo {pages} {e2300860120} (\bibinfo {year} {2023})}\BibitemShut {NoStop}%
\bibitem [{\citenamefont {Yu}\ \emph {et~al.}(2024)\citenamefont {Yu}, \citenamefont {Xue}, \citenamefont {Guo}, \citenamefont {Chan}, \citenamefont {Terh}, \citenamefont {Soci}, \citenamefont {Zhang},\ and\ \citenamefont {Chong}}]{yu2024dirac}%
  \BibitemOpen
  \bibfield  {author} {\bibinfo {author} {\bibfnamefont {L.}~\bibnamefont {Yu}}, \bibinfo {author} {\bibfnamefont {H.}~\bibnamefont {Xue}}, \bibinfo {author} {\bibfnamefont {R.}~\bibnamefont {Guo}}, \bibinfo {author} {\bibfnamefont {E.~A.}\ \bibnamefont {Chan}}, \bibinfo {author} {\bibfnamefont {Y.~Y.}\ \bibnamefont {Terh}}, \bibinfo {author} {\bibfnamefont {C.}~\bibnamefont {Soci}}, \bibinfo {author} {\bibfnamefont {B.}~\bibnamefont {Zhang}},\ and\ \bibinfo {author} {\bibfnamefont {Y.}~\bibnamefont {Chong}},\ }\bibfield  {title} {\bibinfo {title} {Dirac mass induced by optical gain and loss},\ }\href@noop {} {\bibfield  {journal} {\bibinfo  {journal} {Nature}\ }\textbf {\bibinfo {volume} {632}},\ \bibinfo {pages} {63} (\bibinfo {year} {2024})}\BibitemShut {NoStop}%
\bibitem [{\citenamefont {Qin}\ \emph {et~al.}(2024)\citenamefont {Qin}, \citenamefont {Ye}, \citenamefont {Wang}, \citenamefont {Zhao}, \citenamefont {Liu}, \citenamefont {Li}, \citenamefont {Hu}, \citenamefont {Liu}, \citenamefont {Wang}, \citenamefont {Longhi} \emph {et~al.}}]{qin2024observation}%
  \BibitemOpen
  \bibfield  {author} {\bibinfo {author} {\bibfnamefont {C.}~\bibnamefont {Qin}}, \bibinfo {author} {\bibfnamefont {H.}~\bibnamefont {Ye}}, \bibinfo {author} {\bibfnamefont {S.}~\bibnamefont {Wang}}, \bibinfo {author} {\bibfnamefont {L.}~\bibnamefont {Zhao}}, \bibinfo {author} {\bibfnamefont {M.}~\bibnamefont {Liu}}, \bibinfo {author} {\bibfnamefont {Y.}~\bibnamefont {Li}}, \bibinfo {author} {\bibfnamefont {X.}~\bibnamefont {Hu}}, \bibinfo {author} {\bibfnamefont {C.}~\bibnamefont {Liu}}, \bibinfo {author} {\bibfnamefont {B.}~\bibnamefont {Wang}}, \bibinfo {author} {\bibfnamefont {S.}~\bibnamefont {Longhi}}, \emph {et~al.},\ }\bibfield  {title} {\bibinfo {title} {Observation of discrete-light temporal refraction by moving potentials with broken galilean invariance},\ }\href@noop {} {\bibfield  {journal} {\bibinfo  {journal} {Nature Communications}\ }\textbf {\bibinfo {volume} {15}},\ \bibinfo {pages} {5444} (\bibinfo {year} {2024})}\BibitemShut {NoStop}%
\bibitem [{\citenamefont {Feis}\ \emph {et~al.}(2025)\citenamefont {Feis}, \citenamefont {Weidemann}, \citenamefont {Sheppard}, \citenamefont {Price},\ and\ \citenamefont {Szameit}}]{feis2025space}%
  \BibitemOpen
  \bibfield  {author} {\bibinfo {author} {\bibfnamefont {J.}~\bibnamefont {Feis}}, \bibinfo {author} {\bibfnamefont {S.}~\bibnamefont {Weidemann}}, \bibinfo {author} {\bibfnamefont {T.}~\bibnamefont {Sheppard}}, \bibinfo {author} {\bibfnamefont {H.~M.}\ \bibnamefont {Price}},\ and\ \bibinfo {author} {\bibfnamefont {A.}~\bibnamefont {Szameit}},\ }\bibfield  {title} {\bibinfo {title} {Space-time-topological events in photonic quantum walks},\ }\href@noop {} {\bibfield  {journal} {\bibinfo  {journal} {Nature Photonics}\ }\textbf {\bibinfo {volume} {19}},\ \bibinfo {pages} {518} (\bibinfo {year} {2025})}\BibitemShut {NoStop}%
\bibitem [{\citenamefont {He}\ \emph {et~al.}(2025)\citenamefont {He}, \citenamefont {Dong}, \citenamefont {Li}, \citenamefont {Yu}, \citenamefont {Wu}, \citenamefont {Chen},\ and\ \citenamefont {Yuan}}]{he2025observing}%
  \BibitemOpen
  \bibfield  {author} {\bibinfo {author} {\bibfnamefont {Y.}~\bibnamefont {He}}, \bibinfo {author} {\bibfnamefont {Z.}~\bibnamefont {Dong}}, \bibinfo {author} {\bibfnamefont {G.}~\bibnamefont {Li}}, \bibinfo {author} {\bibfnamefont {P.}~\bibnamefont {Yu}}, \bibinfo {author} {\bibfnamefont {X.}~\bibnamefont {Wu}}, \bibinfo {author} {\bibfnamefont {X.}~\bibnamefont {Chen}},\ and\ \bibinfo {author} {\bibfnamefont {L.}~\bibnamefont {Yuan}},\ }\bibfield  {title} {\bibinfo {title} {Observing momentum conservation at temporal interfaces in synthetic frequency dimension},\ }\href@noop {} {\bibfield  {journal} {\bibinfo  {journal} {Science Advances}\ }\textbf {\bibinfo {volume} {11}},\ \bibinfo {pages} {eadz5445} (\bibinfo {year} {2025})}\BibitemShut {NoStop}%
\end{thebibliography}
\end{document}